\documentclass[a4paper,10pt]{article}
\usepackage[utf8]{inputenc} 
\usepackage[T1]{fontenc}    
\usepackage{hyperref}       
\usepackage{url}            
\usepackage{booktabs}       
\usepackage{amsfonts}       
\usepackage{nicefrac}       
\usepackage{microtype}      
\usepackage{lipsum}
\usepackage{graphicx}
\usepackage{authblk}
\usepackage[backend=biber,style=numeric,maxnames=3,minnames=1,sorting=none]{biblatex}
\graphicspath{ {./figures/} }

\title{Multi-modal Representation Learning Enables Accurate Protein Function Prediction in Low-Data Setting}
\author[1,2]{Serbulent Ünsal\thanks{serbulentu@gmail.com}}
\author[1]{Sinem Özdemir}
\author[3]{Bünyamin Kasap}
\author[1]{M. Erşan Kalaycı}
\author[1]{Kemal Turhan}
\author[4,5]{Tunca Doğan\thanks{tuncadogan@gmail.com}}
\author[2]{Aybar C. Acar\thanks{acacar@metu.edu.tr}}

\affil[1]{\small Department of Biostatistics and Medical Informatics, Faculty of Medicine, Graduate School of Health Sciences, Karadeniz Technical University, Trabzon, Turkiye} 
\affil[2]{\small Cancer Systems Biology Laboratory (KanSiL), Graduate School of Informatics, Middle East Technical University, Ankara, Turkiye}
\affil[3]{\small Health Sciences University Trabzon Kanuni Training and Research Hospital, Medical Microbiology Laboratory, Trabzon, Turkiye}
\affil[4]{\small Biological Data Science Lab, Dept. of Computer Engineering, Department of Computer Engineering, Hacettepe University, Ankara, Turkiye}
\affil[5]{\small Dept. of Bioinformatics, Graduate School of Health Sciences, Hacettepe University, Ankara, Turkiye}

\date{}
\begin{document}
\maketitle
\begin{abstract}
    In this study, we propose HOPER (HOlistic ProtEin Representation), a novel
    multimodal learning framework designed to enhance protein function
    prediction (PFP) in low-data settings. The challenge of predicting protein
    functions is compounded by the limited availability of labeled data.
    Traditional machine learning models already struggle in such cases, and
    while deep learning models excel with abundant data, they also face
    difficulties when data is scarce. HOPER addresses this issue by integrating
    three distinct modalities—protein sequences, biomedical text, and
    protein-protein interaction (PPI) networks—to create a comprehensive protein
    representation. The model utilizes autoencoders to generate holistic
    embeddings, which are then employed for PFP tasks using transfer learning.
    HOPER outperforms existing methods on a benchmark dataset across all Gene
    Ontology categories, i.e., molecular function, biological process, and
    cellular component. Additionally, we demonstrate its practical utility by
    identifying new immune-escape proteins in lung adenocarcinoma, offering
    insights into potential therapeutic targets. Our results highlight the
    effectiveness of multimodal representation learning for overcoming data
    limitations in biological research, potentially enabling more accurate and
    scalable protein function prediction. HOPER source code and datasets are
    available at
    \href{https://github.com/kansil/HOPER}{https://github.com/kansil/HOPER}
\end{abstract}


\section{Introduction}

Proteins are macromolecules that are the building blocks and essential machinery
of life. However, knowledge about the functional properties of proteins is still
limited. In particular, the UniProt knowledgebase, the largest information hub
for proteins, has approximately 250 million protein records; however, only 0.3\%
of them are manually reviewed. Manual annotation of protein function requires
wet lab experiments and curation of results by human experts. This is a slow and
high-cost process. Especially in the last decade, rigorous efforts have been
made to annotate proteins with automated systems (a.k.a protein function
prediction -- PFP)~\cite{Friedberg2006-oh,Shehu2016-dm,Makrodimitris2020-vb}.
Machine learning (ML) based protein function prediction has two major issues.
First, there is a need for data preprocessing, since conventional ML methods
rely on manually extracted features (e.g., physicochemical properties of
proteins). Second, the predictive accuracy of these methods is still not
sufficient, as indicated by the CAFA (Critical Assessment of Functional
Annotation) challenge, which is a well-known and periodic PFP
competition~\cite{Zhou2019-iw} based on Gene Ontology (GO)
annotations~\cite{Ashburner2000-fo,Gene_Ontology_Consortium2021-rz} for
molecular function (MF), biological process (BP) and cellular component (CC)
categories. Consequently, PFP is still an open problem.

Deep learning techniques, on the other hand, not only display high performance
in biological data modelling but also eliminate the need for manual feature
extraction~\cite{Littmann2021-oj,Yao2021-ke,You2021-fk}. However, one important
problem of deep learning models is requiring extensive labelled training
datasets to achieve high accuracy~\cite{Gaonkar2016-ch,Ng2020-zc}. The lack of
large-scale data hampers the effectiveness of deep learning in PFP, prompting
researchers to explore alternative strategies to address this limitation. Even
though there are a few PFP studies that consider the low-data problem, e.g.,
DEEPred~\cite{Sureyya_Rifaioglu2019-mw}, DeepGOZero~\cite{Kulmanov2022-jk} and
ProTranslator~\cite{Xu2022-pn}, none of them demonstrated adequate
generalisation in practical scenarios. Hence, new methods are needed for
low-data PFP that use integrative approaches. Additionally, a variety of machine
learning algorithms have been proposed to solve the data problem, such as
few-shot learning~\cite{Stanley2021-mn}, generating synthetic
data~\cite{Akbar2022-cs}, one-shot learning using Siamese
network~\cite{Mostavi2021-mx}, data-efficient reinforcement
learning~\cite{Schwarzer2020-cg}. Applications of these approaches include
predicting the effects of mutations~\cite{Meier2021-ny}, acetylation site
prediction~\cite{Peiran2021-af}, predicting kinase-phosphosite
associations~\cite{Deznabi2020-yi}, protein localisation
prediction~\cite{Khwaja2022-rx}, and protein design~\cite{Liu2021-yc}.
Nevertheless, these solutions are mostly problem-specific; hence, for each task,
new models should be trained and fine-tuned from scratch. This process is costly
in terms of time and funds.

An alternative solution for the problem of large data requirements is
representation learning. In this approach, a model is pre-trained using a large
dataset of unlabeled samples. Thus, a generalised high-dimensional
representation of the input data is learned. This pre-trained model can be used
in subsequent prediction tasks~\cite{Iuchi2021-uh} by either fine-tuning the
original model or training a new supervised model that utilises the
representations/embeddings produced by the pre-trained model as input. These
representations can capture various critical biochemical and functional features
of proteins. As an example, Vig et. al.~\cite{Vig2022-kr} investigated the
transformer architecture~\cite{Vaswani2017-pi} in this context and found out
that attention layers can capture valuable information on structure, contact
points, binding sites and evolutionary information. These findings were
supported by others, who added function prediction to this list through deep
mutational scanning data~\cite{Bepler2021-yc} and fitness
landscapes~\cite{Wittmann2021-ka}.

Most protein representation learning (PRL) models solely use protein sequence
data as input, which is information-rich; however, part of this knowledge is
implicit, which is difficult to extract. Other data types, such as the natural
language-based text from the biomedical literature and molecular relationships
(e.g., protein-protein interactions), can explicitly contain protein-related
information. Therefore, leveraging multiple data types has the potential to
improve protein learning.

Multimodal representation learning, defined as the extraction of knowledge from
a combination of data modalities, is a promising approach utilised in various
fields of data science~\cite{Guo2019-sf}. Techniques such as early and late
fusion~\cite{Baltrusaitis2019-rv}, multiplicative
interactions~\cite{Jayakumar2020-va}, multimodal gated units~\cite{Wang2019-fs},
temporal attention~\cite{Tsai2019-np} (a.k.a multimodal transformers),
architecture search~\cite{Xu2021-rc} have been employed to yield integrative
learning with multiple modalities in the literature. Additionally, alternative
training procedures, i.e., gradient blending and
regularisation~\cite{Scheunemann2019-id} by maximising functional
entropies~\cite{Gat2020-qk}, have been used. An evaluation of these approaches
can be found in the MultiBench study~\cite{Liang2021-rx}.

Multimodal learning has also been exploited in the framework of biological
sequence analysis, such as the prediction of (i) long noncoding RNA-Protein
interactions with capsule networks~\cite{Li2021-yd}, (ii) long noncoding
RNA-miRNA interactions~\cite{Hu2020-wy}, (iii) protein ubiquitination
sites~\cite{He2018-ep}, (iv) protein-protein
interactions~\cite{Zhang2019-un,Xue2021-dd}, (v) moonlighting
proteins~\cite{Li2021-ll}, and (vi) small molecules of novel scaffolds sharing
similar target biological activities~\cite{Zheng2021-gu}.

\begin{figure}
    \centering
    \includegraphics[width=\textwidth]{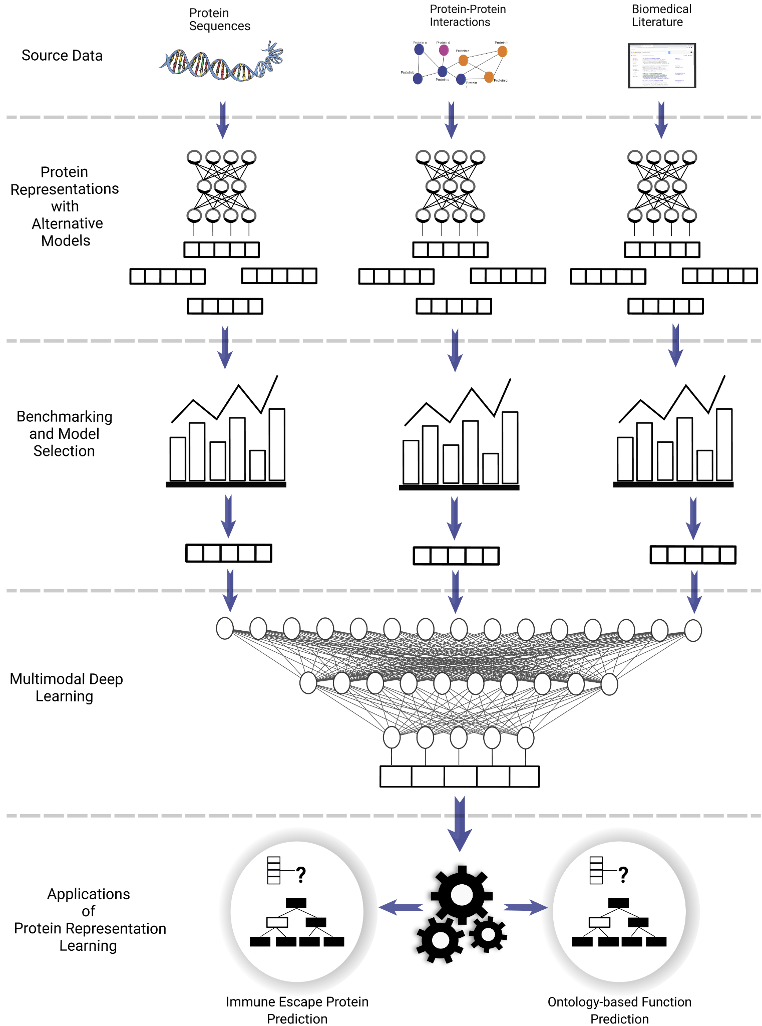}

    \caption{The overview of the study. \small We first generated protein
     representations (embeddings) independently using three different modalities
     (i.e., protein sequence, protein-protein interaction, and text). Then, we
     benchmarked them to find the best-performing representation model for each
     modality in the context of protein function prediction in low training data
     setting. After that, we constructed multimodal learning models that take
     representations of independent modalities as input and produce a holistic
     embedding by leveraging their relationships. Finally, as a use-case study,
     we predicted new tumor immune-escape proteins in the lung adenocarcinoma
     disease using our model and discussed the results.}
    \label{fig:overview}
\end{figure}

In this study, we proposed a holistic protein representation learning model,
HOPER, using multimodal learning to perform high-performance PFP even with low
training data. The outline of our study is given in Figure \ref{fig:overview}.
For this, we first created representation vectors using (i) amino acid
sequences, (ii) protein-based text from the literature, and (iii)
protein-protein interaction data types. We then yielded integrative (multimodal)
learning over those three modalities using autoencoders. The amino acid
sequences of proteins have already been the main type of input data for most of
the protein representation learning methods in the literature. The rationale
behind incorporating protein-protein interactions into our model is the
assumption that interacting proteins are likely to act in the same biological
process and cellular component. Apart from that, the reason for utilising
textual data from the scientific literature is that it contains rich knowledge
directly refined from experimental (wet-lab) results. Such data may include a
comprehensive semantic context for the given protein.

Additionally, we showed that once the sequence is enriched with text and PPI
data, this associative knowledge can be used to develop high-performing
sequence-based low-data PFP models via transfer learning when text and PPI data
are unavailable. This model was developed mainly to learn the relationship
between sequence and other modalities. Our HOPER models were tested on a PFP
benchmark (using the GO categories of MF, BP, and CC), which was explicitly
designed to evaluate protein representation learning methods under different
training sample sizes (low, middle, high) and function specificity levels
(specific, normal, shallow). Finally, as a use case, we employed HOPER to
identify new immune-escape proteins in the lung adenocarcinoma disease.

\section{Results}

Our multimodal protein representation model, HOPER (Holistic Protein
Representation), uses sequence, text, and PPI as data modalities. Here, we
investigated the performance of HOPER and its constituents on the problem of
Gene Ontology-based protein function prediction (PFP).  We developed and tested
a variety of models to combine these modalities, especially using multimodal
learning approaches.

We utilised our previously constructed benchmarking framework, PROBE (Protein
RepresentatiOn BEnchmark), for PFP performance evaluation and
comparison~\cite{Unsal2022-jz}. PROBE was constructed to identify how protein
representation models performed on PFP problems under different function
specificities and different training dataset sizes. We carefully curated nine
different protein function datasets considering sample counts (i.e., low,
middle, and high), and GO term specificities (i.e., specific, normal, shallow).
We selected groups of GO terms (mostly composed of 5 terms for each group) based
on combinations of these two parameters. The benchmark framework covers the
molecular function (MF), biological process (BP), and cellular component (CC)
aspects of GO. For instance, the benchmark dataset referred to as
``MF\_Low\_Shallow'' includes 5 MF GO terms that are closer to the root of the
GO MF tree (i.e., ``shallow'' in terms of term specificity), each of which has 5
to 30 protein annotations (i.e., ``low'' in terms of the number of samples).
Details about the other datasets and the included GO terms can be found in the
Methods section.

Below, we first explain the results of the PFP analyses done independently on
three different types of data modalities. Then, we elaborate on the multi-modal
learning approaches we utilised to yield holistic learning over the sequence,
PPI, and text modalities and their subsequent PFP benchmark performance outcome.

\subsection{Evaluation of sequence-based representation models}

We evaluated 23 protein representations using the PROBE
benchmark~\cite{Unsal2022-jz}. We employed a linear classifier to decouple the
classification's final performance from the representation model. We utilized a
linear boundary classifier to ensure that the test results accurately reflect
the models' ability to learn meaningful high-dimensional representations of the
protein space. The ideal representation should ensure that the proteins'
functional properties are linearly separable.

\begin{figure}[htbp]
    \centering

    \includegraphics[width=\textwidth]{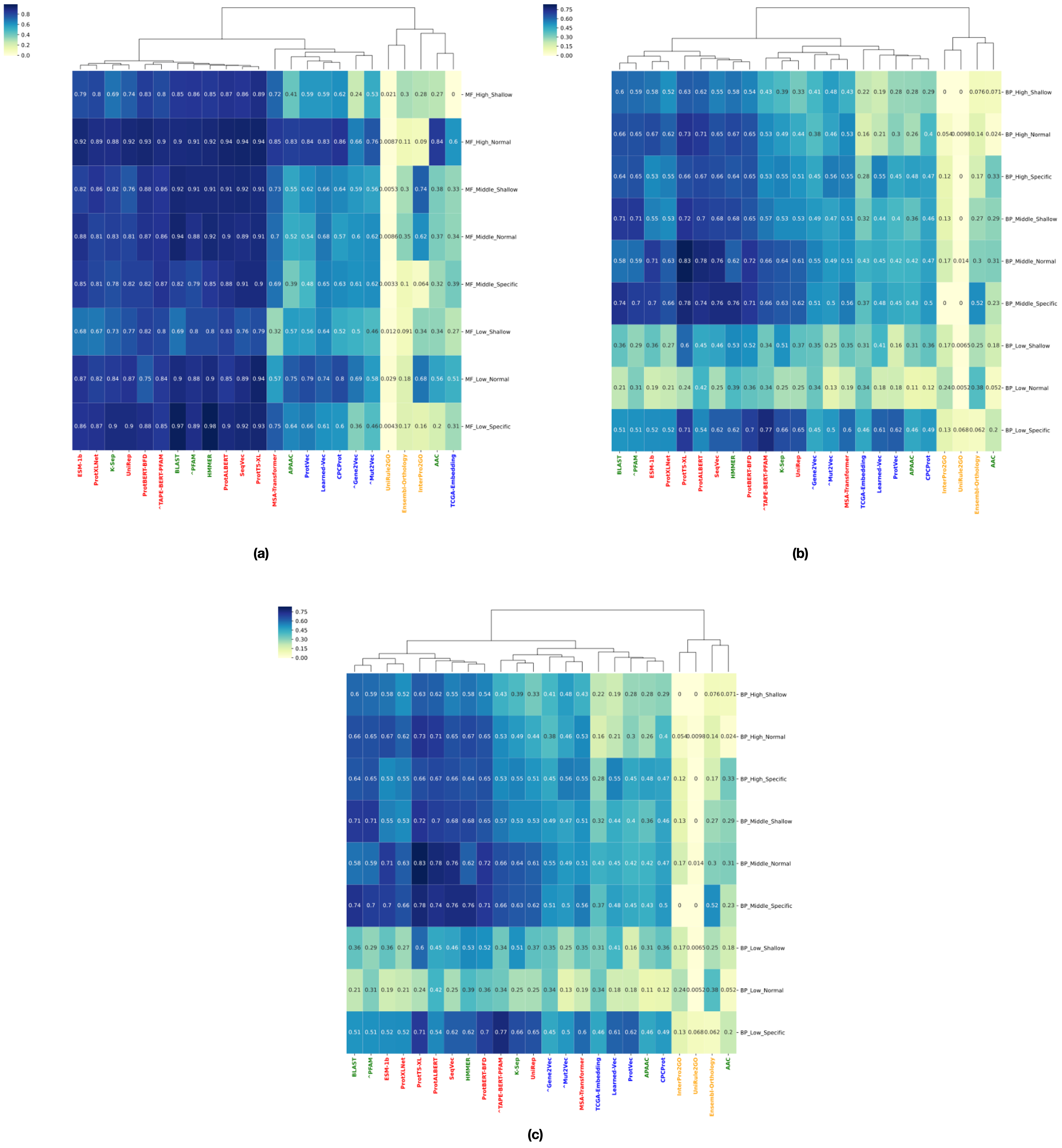}
    
    \caption{Ontology-based protein function prediction benchmark results. Heat
    maps indicating the clustered performance results (weighted F1-scores) of
    protein representation methods in ontology-based PFP benchmark in terms of
    GO categories of \textbf{(a)} molecular function, \textbf{(b)} biological
    process and \textbf{(c)} cellular component. The colours indicate groups of
    models (yellow, rule-based annotation methods; green, classical
    representations; blue, small-scale learned representations; red, large-scale
    learned representations).}\label{fig:benchmark}
\end{figure}

Figure \ref{fig:benchmark} and Supplementary Table S1a display the GO prediction
performance results of 23 protein representations. In the benchmark, learned
representation models generally outperformed classical methods, especially in
predicting Molecular Function (MF) GO terms. Performance was lower for Cellular
Component (CC) and Biological Process (BP) predictions, likely due to the
reliance on sequence data, which isn't a strong indicator for these tasks.
Prediction accuracy declined with fewer annotated proteins, particularly for CC
terms, but there was no clear difference based on term specificity. However,
challenges remain in predicting specific GO terms due to limited annotation. Our
evaluation of various PFP methods for predicting Gene Ontology (GO) terms
revealed that some of the protein language models (e.g., ProtT5-XL, ProtALBERT,
SeqVec, ProtBERT, etc.) outperformed classical methods (e.g., BLAST, KSEP, etc.)
significantly. We observed that the success rate in predicting GO terms
decreased with fewer annotated proteins (i.e., smaller training datasets),
especially for specific/informative terms. Despite this, we found that some of
the top-performing methods demonstrated consistent performance across virtually
all GO groups, with ProtT5-XL~\cite{Elnaggar2022-ja} emerging as the best
performer in all three categories (CC, BP, and MF), followed closely by
ProtALBERT, SeqVec, ProtBERT-BFD, and HMMER.

\subsection{Evaluation of protein-protein interaction-based representation models}

For this analysis, the protein-protein interaction (PPI) data is encoded as a
graph. We selected two widely used graph representation methods with reference
code implementations, High-Order Proximity Preserved Embedding
HOPE~\cite{Ou2016-er} and Node2Vec~\cite{Grover2016-bo}. We applied these
methods to the PPI graphs to calculate node (i.e., protein) representation
vectors. We explored various hyperparameters for these methods and visualized
the best performing 10 parameter combinations for each. Selected hyperparameters
for the HOPE method are ``d'' and ``beta''. Likewise, ``p'', ``q'' and ``d'' are
tuned for Node2Vec. We explain these hyperparameters in the Methods section.

The HOPE method has two aims: preserving high-order proximity and capturing
asymmetric transitivity. On the other hand, Node2Vec uses a random walk and
word2vec~\cite{Mccormick2016-uk} to compute node embeddings in graphs. We used
the IntAct database (accessed on September 6, 2020) to acquire PPI data and used
only the interactions between human proteins. The final network includes 16,435
nodes (proteins) and 241,833 edges ( protein-protein interactions).

\begin{figure}[htbp]
    \centering

    \includegraphics[width=\textwidth]{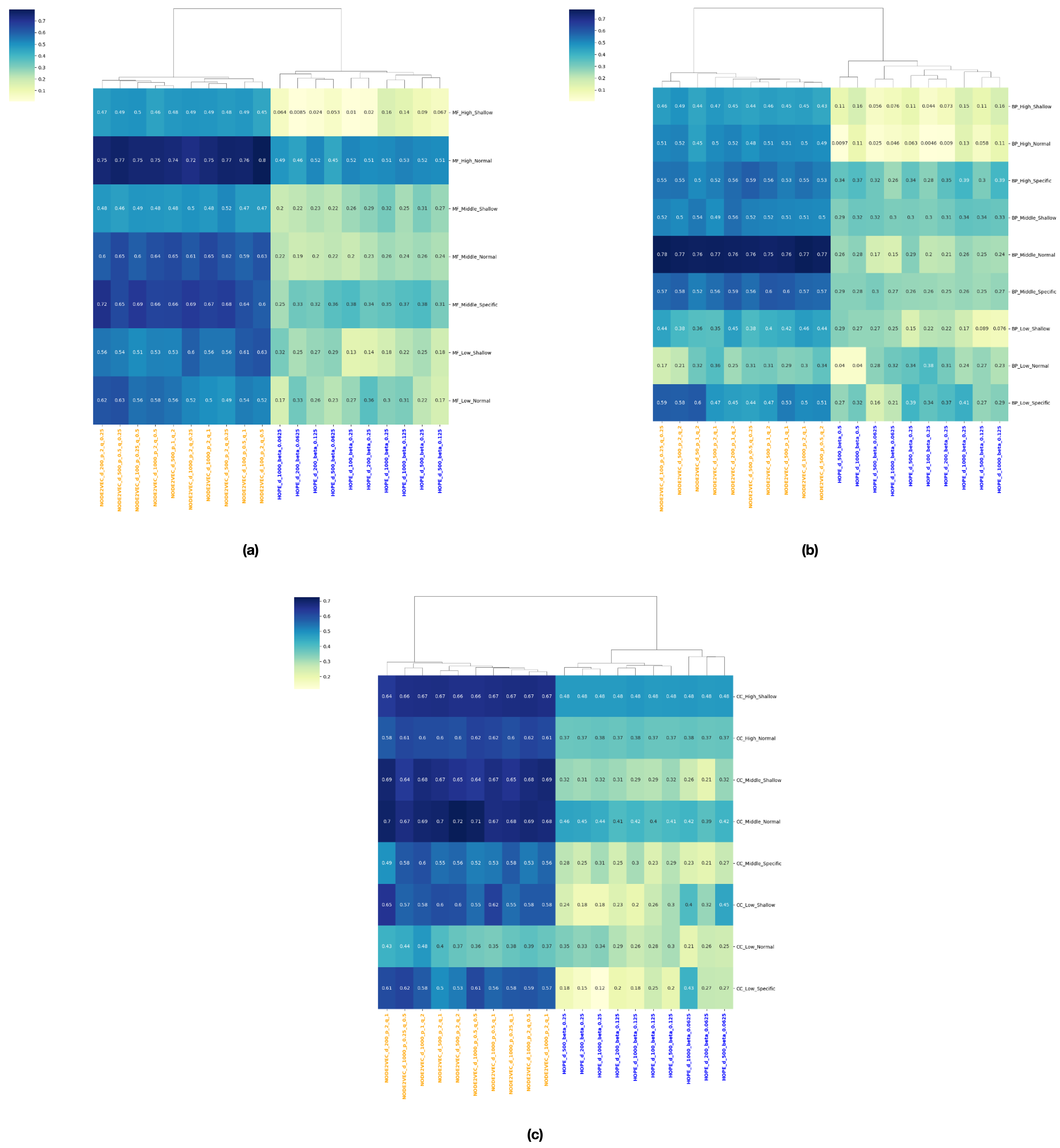}
    
    \caption{Protein function prediction performances of PPI representations
    based on mean F1-weighted scores. Multiple models that utilize Node2Vec and
    HOPE algorithms are shown with different hyperparameter selections (p and q
    for Node2Vec and beta for HOPE). Tests were conducted on the low data
    setting (i.e., MF, BP, and CC GO terms with low number of annotated
    proteins) for; \textbf{a} MF; \textbf{b} BP; and \textbf{c} CC
    terms.}\label{fig:PPI_perf}
\end{figure}

Figure \ref{fig:PPI_perf} and Supplementary Table S1b show that Node2Vec was
more accurate than HOPE on the PFP problem. A possible reason is a mismatch
between the objectives of the HOPE model and the particulars of our use-case:
The first objective of the HOPE is high-order proximity. However, in PPI
networks, low-order proximities are more indicative of functional relationships.
Similarly, HOPE's second objective, capturing the asymmetric transitivity, might
not apply to our data since our graph is undirected.

Another important observation is the relation between hyper-parameters and model
performance. As shown in Figure \ref{fig:PPI_perf} and Supplementary Table S1b,
model performance is sensitive to hyperparameters.

\subsection{Evaluation of text-based protein representation models}

Text data, obtained either from curation or directly from the literature, can be
considered one of the most refined data sources to acquire knowledge on protein
functions. However, this information is highly distributed and heterogeneous. We
aimed to incorporate this knowledge as a protein representation. To accomplish
this, we used two data sources; curated UniProt text from flat files (for 20,365
human protein records in total), and scientific literature (in the form of
article abstracts) for selected human proteins using PubMed records of the
articles cited in the curated text of the protein of interest in UniProtKB
(17.639 protein records in total). We obtained text data from the ``Comments
section'' for each human protein entry.

We evaluated alternative methods to calculate textual representations of the
proteins in our dataset. First, we apply the term frequency--inverse document
frequency TF-IDF~\cite{Aizawa2003-hk} to the text data which is explained in the
methods section. The size of the original TF-IDF vectors was determined
according to the number of unique terms (mostly words) in the corpus. However,
the resulting vectors were large (54428 dims for UniProtKB retrieved text and
178451 dims for UniProtKB\&PubMed retrieved text) and processing them directly
would have been both computationally intractable and might have caused problems
due to the curse of dimensionality. To mitigate this problem, we applied
principal components analysis (PCA) and constructed vectors composed of 256, 512
and 1024 principal components as ``reduced TF-IDF vectors''. We called this
method TF-IDF\_PCA.

Afterwards, we used pre-trained text representation models to generate
text-based protein representations. We first evaluated the
BioWordVec~\cite{Zhang2019-ya} model which uses the classical word2vec algorithm
and is pre-trained on the PubMed text corpus and MIMIC III data containing
2,324,849 distinct words in total. The number of dimensions in BioWordVec
vectors is 200\. We also utilized BioSentVec~\cite{Chen2019-qo} . BioSentVec
uses the sent2vec~\cite{Moghadasi2020-se} model to compute 700-dimensional
sentence embeddings and is trained on the same dataset as BioWordVec. Similarly,
we utilized the BioBERT~\cite{Lee2020-fu} model. The BioBERT model is a
BERT~\cite{Devlin2018-ag} based model which is trained on biomedical and
clinical corpus, more details can be found in the methods section. The BioBERT,
BioWordVec, and BioSentVec produce representation vectors per word. We
max-pooled or averaged these vectors for all words in the text data per protein.

\begin{figure}[htbp]
    \centering

    \includegraphics[width=\textwidth]{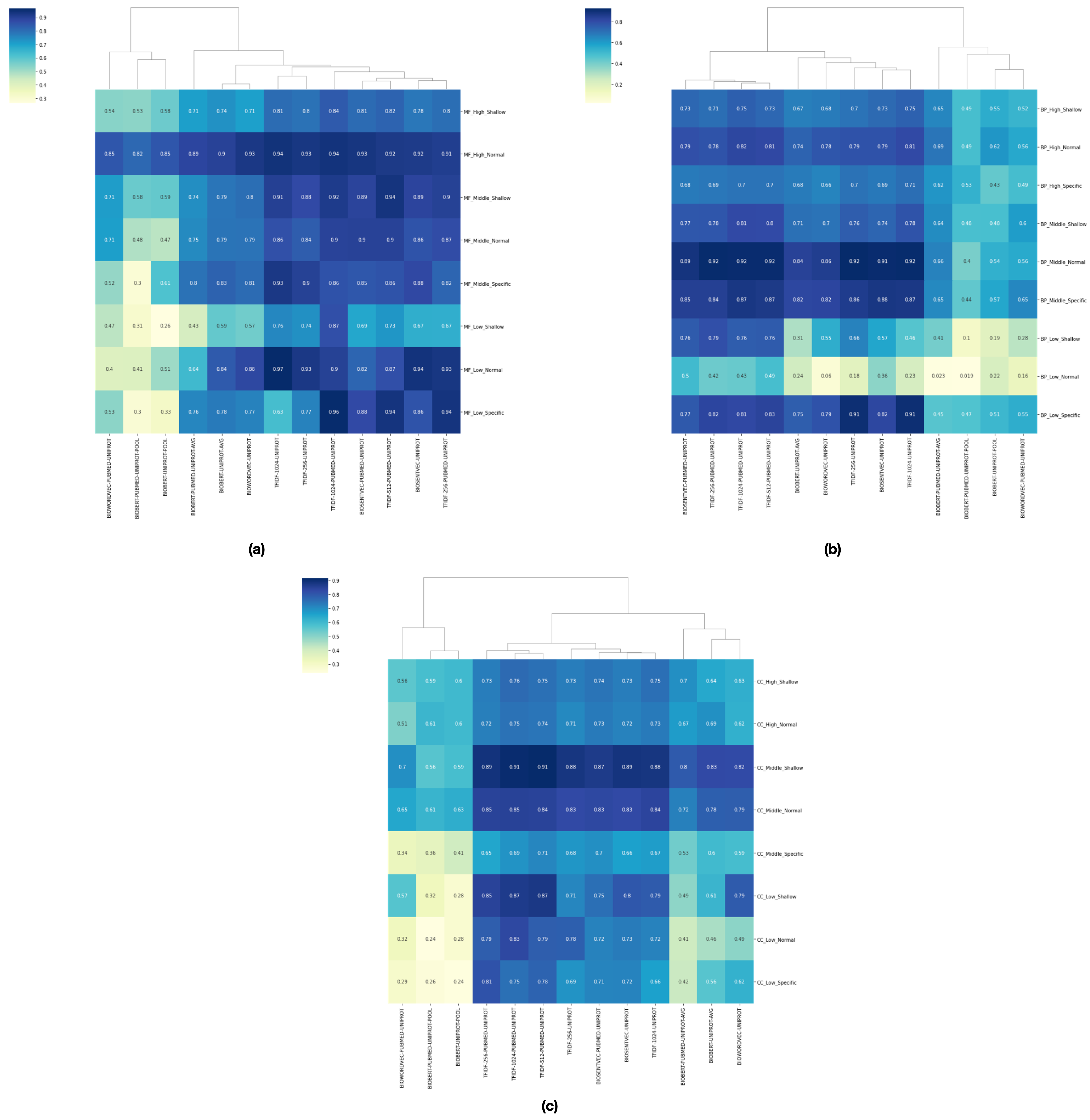}
    
    \caption{Protein function prediction performances of text-based
    representations based on mean F1-weighted scores. Multiple models that
    utilize TF-IDF-PCA, BioSentVec, and BioBERT algorithms are shown with
    different hyperparameter and data source selections. Tests were conducted on
    the low data setting (i.e., MF, BP, and CC GO terms with low number of
    annotated proteins) for; \textbf{a} MF; \textbf{b} BP; and \textbf{c} CC
    terms.}\label{fig:text_perf}
\end{figure}

We observed that TF-IDF\_PCA produced the best performance on the PFP problem in
the text-based models. The best performance with TF-IDF\_PCA is achieved when
UniProt and PubMed data are jointly used as the input. Even though TF-IDF\_PCA
is the most simplistic model and showed the top performance, the BioBERT, the
most sophisticated and extensive sized model compared to other models we
evaluated (according to number parameters), was the worst performed overall
(Figure \ref{fig:text_perf} \& Supplementary Table S1c). In the natural language
processing (NLP) domain, these approaches were successful, especially on data
composed of short sentences~\cite{Wang2018-qo}. However, models specialized in
processing longer sequences are required here such as
Longformer~\cite{Beltagy2020-yo} and Reformer~\cite{Kitaev2020-wc}.
Unfortunately, none of the available long-sequence models are pre-trained in
biomedical texts. On sub-category-based performance, all models underperformed
in the ``low-data'' category. For other categories, no clear pattern was
observed.

\subsection{Generating a holistic protein representation using autoencoders}

Currently, representations that aim to integrate multiple types of protein
features (i.e multimodal models) are limited especially for
PFP~\cite{Giri2021-ch,Gligorijevic2018-kh,Li2021-ll}. Moreover, according to the
best of our knowledge, none of these models specialized in producing accurate
predictions in a low-data setting. Another problem in the application of
multimodal approaches is the scarcity of data, such as experimentally verified
protein-protein interactions and curated biomedical texts. For example, in our
study, we analyzed the human proteome, which is one of the most extensively
studied protein collections in the literature. Here, a total of 20,421 proteins
were found to possess sequence data, whereas 16,435 proteins exhibited PPI.
Additionally, 17,639 proteins were annotated with textual information. Notably,
14,941 proteins were found to possess all three data types, thereby providing a
multi-dimensional perspective of these proteins. Here, we aimed to circumvent
these problems and develop multimodal models that can learn relationships
between different types of protein data (i.e., text, PPI, and sequence) and
generate holistic protein representations.

\begin{figure}[htbp]
    \centering

    \includegraphics[width=\textwidth]{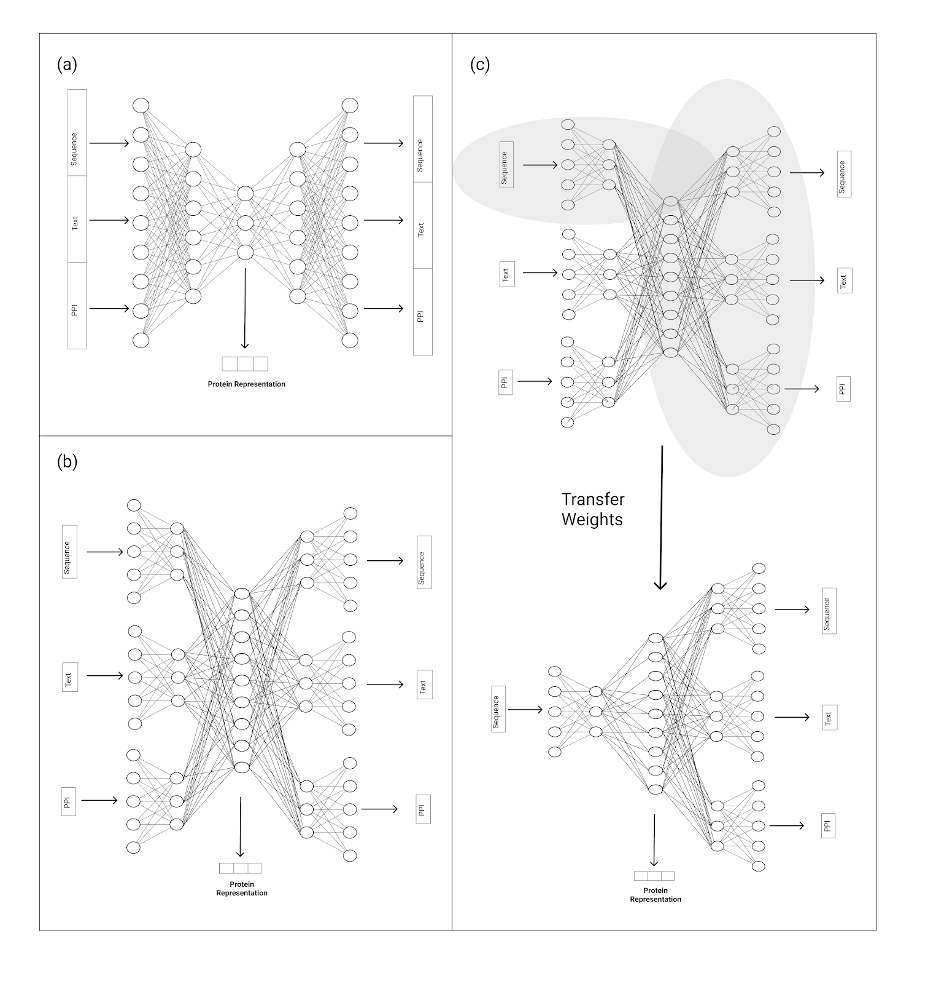}
    
    \caption{Construction of multimodal representations using autoencoder
    models; \textbf{a} simple autoencoder (SimpleAE), \textbf{b} multimodal
    autoencoder (MultiModelAE), and \textbf{c} transfer learning-based
    sequence-input multimodal autoencoder (TransferAE).}\label{fig:multimodal}
\end{figure}

The first approach we considered for leveraging multiple modalities is a
concatenation of representations generated by text, PPI, and sequence. We have
created all subsets of these representations. The second approach we applied is
training a simple autoencoder (Simple-AE) which takes a concentrated
representation consisting of text, PPI, and sequence and aims to reconstruct the
input representation (Figure \ref{fig:multimodal}a). The third approach we used
was a multimodal autoencoder (MultimodalAE), which takes input from each
modality separately, encodes it using its shared middle layer, and tries to
reconstruct each representation separately (Figure \ref{fig:multimodal}b). The
fourth approach we applied was a transfer learning-based multimodal sequence
autoencoder. We named this model TransferAE (Figure \ref{fig:multimodal}c).  At
the end of their training procedures, we used the bottleneck layer of these
autoencoders as our protein representations and utilized them in our PFP models.
We compared numerous representations with varying bottleneck sizes to optimize
PFP performance. We trained a simple linear support vector machine (SVM) model
as the protein function predictor.

\begin{figure}[htbp]
    \centering

    \includegraphics[width=\textwidth]{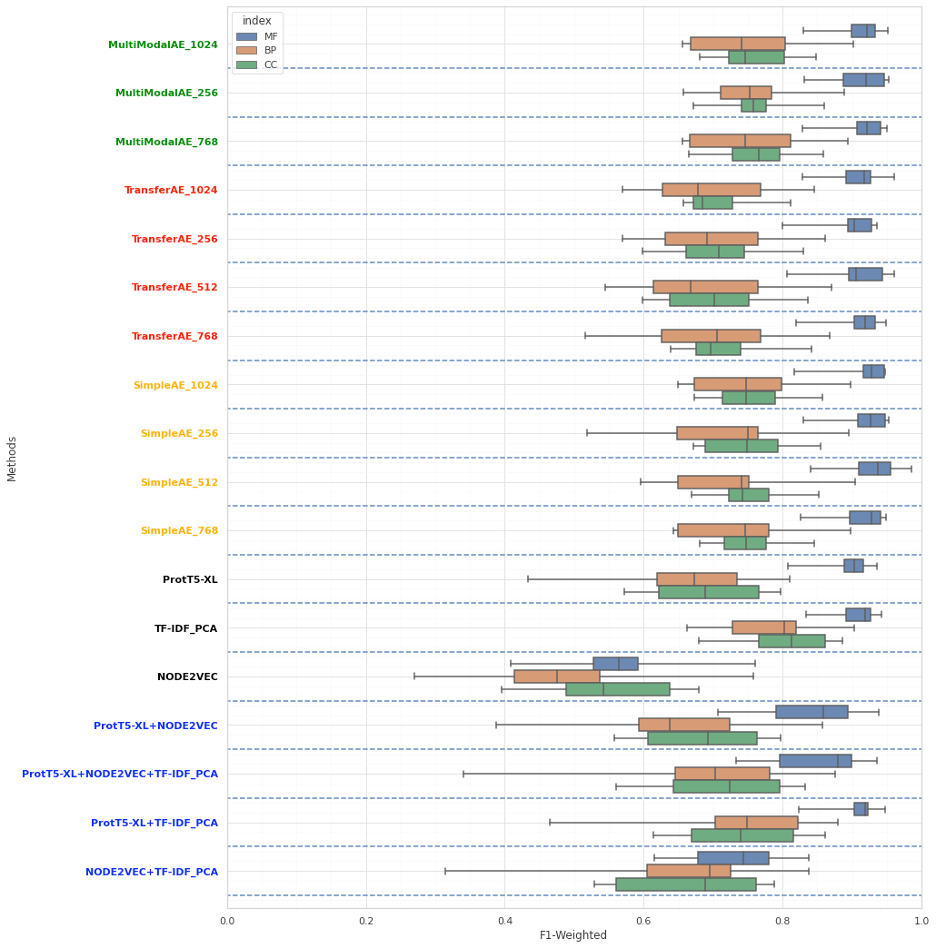}
    
    \caption{Overall performance evaluation of multimodal protein
    representations in protein function prediction, based on mean F1-weighted
    scores. The study showcases various models; black: the best performing
    single modality, blue: fused (FusedRep), yellow: simple autoencoder
    (SimpleAE), green: multimodal autoencoder (MultiModalAE), and red: transfer
    learning-based autoencoder (TransferAE) representation models with varying
    embedding dimension sizes. The tests were conducted on the Gene Ontology
    (GO) categories of molecular function -- MF (blue bars), biological process
    -- BP (orange bars), and cellular component -- CC (green bars), considering
    all dataset size and term specificity-based groups.}\label{fig:overall_perf}
\end{figure}

Figure \ref{fig:overall_perf} and Supplementary Table S2 show that the simple
autoencoder model (Simple\_AE\_512) performed best in the MF category. For the
BP and CC categories, TF-IDF\_PCA produced the best performance. The difference
between TF-IDF\_PCA and simple autoencoder models was not significant. These
results indicate that none of the multimodal models were able to attain a
synergistic effect to outperform text. One other reason might be a suspected
data leak from text to Gene Ontology terms, which we will elaborate on in the
discussion section.

When comparing the best sequence-based model (ProtT5-XL) with TransferAE\_768,
the latter demonstrates significantly improved performance in the low category,
with weighted F1-scores of 0.92 vs. 0.85 for MF, 0.68 vs. 0.55 for BP, and 0.68
vs. 0.61 for CC. Detailed results are provided in Supplementary Table S2. Here,
our primary goals were, (i) developing a multimodal model for functions that
have a low-number of samples (i.e., proteins annotated to the function of
interest), and (ii) have the capability to assign functions to understudied
proteins (e.g., proteins for which only sequence data is available). Therefore,
these goals were satisfied with the TransferAE model.

Finally, we visualized the best-performing representations (i.e.,
Simple\-AE\_\-512 for MF and ProtT5-XL + Node2Vec + TF-IDF\_PCA for BP and CC)
on the 2-D plane for each protein function dataset. We employed the t-SNE
algorithm~\cite{Van_der_Maaten2008-cu} for dimensionality reduction. The
scikit-learn (v1.2.1) manifold TSNE module is used with default parameters.
Resulting density plots showed how well each term could be separated by the
representation at hand (Figure S1).

\subsection{Case Study: Immune Escape Protein Prediction}

As a case study, we aimed to predict new immune escape proteins in lung
adenocarcinomas, which is the cancer type with the highest mortality
rate~\cite{noauthor_2015-kr}. Immune escape can be defined as a tumor's ability
to evade immune elimination~\cite{Beatty2015-lv}. Although this mechanism has
critical importance in cancer research, the current knowledge of it is not
sufficient to develop a broad spectrum of therapeutics. Related to this, the
current number of proteins that are known to take part in immune escape is
limited (\~28 proteins considering all types of cancer). Therefore, there is a
current requirement to discover other proteins that take part in immune escape.
This problem can be considered a good example of the critical requirement to
develop a prediction model with a low amount of training data.

To accomplish this task, we conducted the following steps; (i) We obtained
differentially expressed genes for the lung adenocarcinoma using TCGA-LUAD
dataset RNA-Seq data~\cite{The_Cancer_Genome_Atlas_Research_Network2012-bw},
which constituted our positive dataset;  (ii) created the positive and negative
samples datasets; (iii) constructed our main immune espace prediction model by
taking the multimodal representations from our best performing AE-based
representation model for low-data (which was found in the previous analysis
shown in Supplementary Table S2) and trained FC-NN using these vectors as input
and (v) analyzed results via literature search and existing functional
annotations in source databases.

The detailed process is explained below for each step;

\paragraph{(i)} We acquired TCGA-LUAD RNA-Seq data as stated above and
calculated fold changes using read counts. We used the General Linear Model
(GLM) and fit read counts to a negative binomial distribution before testing
them to find significance scores for the calculated fold changes. We applied the
glmLRT function from the EdgeR package for these steps. Hence, we obtained log
fold-change (logFC) scores and respective false discovery rates (FDRs) at the
end of this process. 

In the end, we found 1434 differentially expressed genes, of which 329 were low,
and 1105 were highly-expressed in lung adenocarcinoma tumors compared to the
matched normal lung tissue (Hoper Differentially Expressed LUAD Genes)
(Supplementary File 1).

\paragraph{(ii)} We created our own positive and negative training datasets. We
curated 24 proteins from the literature that is known to take part in the immune
escape process (Supplementary Table S11). Our initial negatives candidate set
consists of human proteins that have GO BP annotations. Using Uniref50 clusters
we selected representative proteins from our candidate set. So, we ensure that
none of the candidates have a sequence similarity of more than 50\% against each
other. Afterwards, we calculate semantic similarity of each protein to the known
immune-escape proteins using Lin similarity~\cite{Lin1998-je} and GO Biological
Process (BP) ontology. Then we create a set from these proteins which have less
than 50\% semantic similarity with the positive set (known immune-escape
proteins), on average. Finally, we chose 150 proteins from this set randomly to
create a dissimilar but diverse set of negative samples (Supplementary Table
S12)\footnote{See PROBE study ``Ontology-based Protein Function Prediction
Benchmark'' subsection under the Methods section for details.}.

\paragraph{(iii)} Using our best autoencoder-based representation (i.e.,
MultiModalAE-256), we trained immune-escape predictors models within a
hyperparameter search. We utilized fully connected feed-forward neural networks
(FC-NN), to predict new immune-escape proteins. We accrued  macro-averaged
precision 0.80, macro-averaged recall 0.72 (Supplementary Table S7). Using this
model, we predicted possible immune-escape proteins among the differentially
expressed gene set. In the end, we positively predicted 121 proteins from 1434
differentially expressed genes/proteins. These proteins were evaluated with a
literature search, and 20 potential LUAD immune-escape therapy targets were
identified (Supplementary Table S4). Below, we discussed two of these proteins
in the framework of a literature survey.

One of the positively predicted proteins, matrix metalloproteinase 9 (MMP9) has
recently been suggested as a potential therapeutic target for cancer
immunotherapy due to its role in tumor escape from anti-angiogenic therapies.
MMP9 plays an important role in the angiogenesis, invasion, and immune escape of
tumors and has been studied in relation to the production of type IV and type V
collagens in the extracellular matrix. This increases the vascularization of
tumors, resulting in increased tumor growth. Additionally, MMP9 increases the
production of Treg cells, which further aids in tumor invasion. MMP9 stimulates
the release of cytokines, which activate the production of Treg cells. The
increased Tregs then suppress the normal immune response and allow the tumor to
escape the body's defenses~\cite{Ohue2019-tf}.  Further investigation of MMP9
revealed that it cleaves the osteopontin isoform and increases the number of
Myeloid-derived suppressor cells (MDSC) in the tumor tissue, suppressing the
immune response and allowing the tumor to escape the effects of immunotherapies.
MMP9 first cleaves the osteopontin isoform, which in turn stimulates the release
of pro-inflammatory cytokines. These cytokines then activate MDSC proliferation,
leading to an increase in their numbers. Hence, MMP9 is a key player in the
angiogenesis, invasion, and immune escape of tumors and could be a potential
drug target for cancer immunotherapy~\cite{Zeng2022-vo}. Here, a critical
observation is our model's prediction of MMP3 also as an immune-escape protein.
Our further investigation showed that MMP3 is an activator of
pro-MMP9~\cite{Fridman2003-pa}. Hence, our model has the capacity to predict
multiple actors of the immune-escape pathways.

Another protein that is predicted by our FC-NN model to play a role in the
immune escape process, tumor necrosis factor ligand superfamily member 11
(TNFSF11 or RANKL), is a cytokine belonging to the TNF alpha cytokine family
that binds to the TNFRSF11A/RANK receptor. RANKL is secreted directly by tumor
tissue to create an immunosuppressive microenvironment in tumors and  binds to
RANK receptors on the surface of dendritic cells. This binding leads to the
activation of a signaling pathway that ultimately results in the production of
TGFB by the dendritic cells. The signaling pathway activated by RANKL involves
several intermediates, including TRAF6, NFKB, and MAP kinases. These
intermediates ultimately activate the Smad family of transcription factors,
which are involved in the regulation of TGFB expression~\cite{Li2008-il}. TGFB
binds to receptors on the surface of T cells, which activates a signaling
pathway that leads to the expression of Foxp3, a transcription factor that is
necessary for the development and function of TREG cells~\cite{Van_Dam2019-cf}.
Hence, once TGFB is produced, it acts on T cells to promote the differentiation
of TREG cells which suppress T Cells and causes immune-escape. Overall, the
pathway from high levels of RANKL to the formation of TREG cells causes immune
suppression and tumor immune-escape as well.

Apart from positively predicted proteins, successfully predicting negative
samples is a critical task. The immune system is wide and complex, involving
numerous proteins (\>1,500~\cite{Dhillon2020-vj}, thus, it can be challenging to
prove that a specific protein is not related to immune system function. We
explored negative examples to elucidate the limited relationships between the
proteins of interest and the immune system. These are the proteins that exist in
our input list (i.e., the ones that were differentially expressed in the lung
adenocarcinoma), but are predicted not to play a role in immune escape by the
FC-NN model. One such protein is AKR1C2, a member of the aldo/keto reductase
superfamily that catalyzes the conversion of aldehydes and ketones to alcohols.
While AKR1C2 is indirectly associated with myeloid-derived suppressor cells
(MDSC) via PGE2, it is not directly involved in immune system
function~\cite{Ahmed2022-nj}. Collagens, including COL1A1, COL3A1, COL5A1,
COL5A2, and COL11A1, are structural proteins that provide strength and stability
to tissues. Although collagens are present throughout the body, they do not
primarily participate in immune system function except for
infiltration~\cite{Geng2021-ok}. A second example, ALB (albumin) is synthesized
in the liver and primarily functions as a transporter for various substances
such as drugs and hormones. While the liver has immune-related functions, ALB's
primary role is not directly related to the immune system~\cite{Wilde2019-pz}. A
third example, CPS1 (carbamoyl phosphate synthetase 1) is an enzyme that
facilitates the removal of excess nitrogen from the body via the urea cycle.
Even though this process is important for health and homeostasis as a whole,
CPS1 does not directly affect the immune system, even though some studies have
looked at the relationship between nitrogen metabolism and the immune system
~\cite{kurmi2020-408}. Finally, CALCA
(\href{https://www.uniprot.org/uniprotkb/P06881/entry}{Calcitonin gene-related
peptide 1}) is a peptide hormone that regulates calcium and phosphate metabolism
in bones. While calcium and phosphate are crucial for immune
function~\cite{He2022-cb,Gryshchenko2021-he}, CALCA itself does not directly
participate in the immune system.

\section{Discussion}

In this study, we aimed to address the challenge of predicting protein functions
with a limited amount of training data by developing multimodal protein
representations. To validate the efficacy of these representations, we created a
benchmarking tool capable of evaluating their performance in the context of
low-data protein function prediction (PFP). Through the examination of various
representation modalities and methods, we demonstrated the usefulness of our
multimodal protein representations. As a case study, we applied these
representations to the task of predicting immune-escape proteins in lung
adenocarcinoma, showcasing their ability to assist in answering research
questions despite the constraints of limited data availability.

\noindent \textbf{Our sequence-based multimodal representation demonstrates superior performance compared to alternative approaches for low-data protein function prediction.}

Our overall benchmarking results revealed that a simple text-based
representation (TF-IDF\_PCA) achieved the best performance for predicting gene
ontology (GO) biological process (BP) and cellular component (CC) terms. This
method was also nearly as successful as the best method for predicting molecular
function (MF) terms. Unexpectedly, text-based representations outperformed
multimodal representations, which may be the result of an indirect data leak.
The wet-lab results extracted from scientific literature are used both to
generate the curator-written text data offered by UniProt and to assign GO
annotations to proteins, which may have caused an unfair advantage to text-based
representations. The only exception to this was the MF category, where the
sequence-based representations performed better, probably due to the direct
relationship between the protein sequence/structure and the molecular function
it performs (e.g., presence of a kinase domain indicating the protein's amino
acid phosphorylation function -- GO:0006468).

A comparison among text-based representation models revealed that, the most
basic approach, term frequency-inverse document frequency (TF-IDF), demonstrated
higher performance than more complex methods such as BioBERT in all
subcategories of low-data PFP including MF, BP, and CC. One potential
explanation for this result is the pooling process used by large language models
like BioBERT~\cite{Lee2020-fu}. These models generate a representation vector of
size N for each word in a text, resulting in a matrix with dimensions L x N for
a text with L words. This matrix must then be pooled into a fixed-size vector of
N dimensions. One common method for pooling is to take the average of all
vectors in the matrix, which can capture the overall meaning of the text, but
may also result in information loss for long texts. Alternative pooling methods
such as max pooling, which selects the maximum values for each dimension in the
matrix, may also result in information loss in a similar manner. There are
examples which TF-IDF or inverse document frequency (IDF) based approaches
produce equal or better results than BERT~\cite{Devlin2018-ag} based solutions.
For example, in a recent study researchers evaluated TF\_IDF and BERT for long
texts and multi-label document classification and observed that performance of
the TF-IDF is better than the BERT model~\cite{Hirschmeier2020-lc}. In another
study the authors showed that for long legal text classification performance of
TF-IDF and BERT is comparable and their approach combines inverse document
frequency (IDF) and PCA performed better than the BERT model~\cite{Chen2022-ua}.
Finally, in another legal text classification study, a combined approach of
TF-IDF and logistic regression showed better performance on multiple NLP tasks
than the DistilBERT~\cite{Sanh2019-wq} model which is a BERT derived but more
enhanced model~\cite{Tuggener2020-zl}.

In this study, our primary objective was to develop a novel protein function
prediction (PFP) representation that is effective when the positive class in the
training set is limited. This scenario, referred to as low-data PFP, presents a
significant challenge for PFP methods. As mentioned above, in our analysis, the
best method in terms of the mean results on all categories (i.e., not only
considering the low data setting) was TF-IDF\_PCA. When evaluating low-data PFP
results, we observed an improvement in performance by SimpleAE and MultiModalAE
over the leading baseline method, TF-IDF\_PCA, for both GO MF and BP categories.
These findings suggest that multimodal methods hold significant promise for
improving PFP accuracy in low-data scenarios.

Although text-based models are generally effective in PFP, a major limitation of
text data is its scarcity, particularly for under-studied proteins. To address
this issue, we developed TransferAE, which can learn relationships between
protein sequences and data generated by other modalities (e.g., protein-protein
interaction and text from literature). When the results were analyzed, we found
that TransferAE performed better than the best sequence-based (single modal)
model ProtT5-XL~\cite{Elnaggar2022-qd}, demonstrating the efficacy of our
approach. Our results show that knowledge from multimodal models can be
effectively transferred to sequence-only models, improving protein function
prediction in low-data scenarios. This suggests that cross-modal pattern
learning enhances performance, making multimodal approaches valuable for
studying novel proteins where additional modality-based data is limited.

\noindent \textbf{Our case study indicates multimodal learning might have applications in research and the clinic.}

In our case study, the objective was to develop a predictive model for
identifying immune-escape proteins in lung adenocarcinoma using the proposed
multimodal machine learning approach. We have used our best multimodal
representation (MultiModalAE-256) for low-data GO BP PFP, and trained
classification models that take this representation as input, using fully
connected neural networks (FC-NNs). Our model was initially provided with a set
of 1,434 potential proteins, from which it identified 121 immune-escape
candidates. Of these, 20 were subsequently validated through literature
analysis, underscoring the success and efficiency of our approach, which
significantly reduces the time required for manual selection. It is worth noting
that the time required for the selection of the 121 immune-escape protein
candidates would be significantly longer if conducted manually. Given the
importance of time in cancer research together with monetary costs of wet-lab
experiments and clinical trials, our models have the potential to expedite the
identification of potential therapeutic targets. By automatizing the process of
candidate gene/protein selection, accurate computational predictions reduces the
burden on human experts.

We validated a few positive and negative predictions of this model from the
literature, which indicated that the model's results have biological relevance.
Nevertheless, it was not possible for us to perform a full literature-based
validation on all of the predictions (similar to the one done on the 121
positive predictions of the baseline model), since manual inspection is both
time- and labor-intensive, especially considering the estimated 1500 proteins
that are involved in the human immune system~\cite{Dhillon2020-vj}. However, it
is imperative to recognize that this lack of validation does not necessarily
signify that our positive predictions are false positives, as they could be
immune-escape proteins that are yet to be discovered or documented. This point
is especially important considering the fact that precision scores of all
models, including SVM, were measured extremely low, due to the high number of
hits that were considered as false positives, which may turn out to be true
positives after a thorough literature search. In a similar fashion, given the
complex nature of the human immune system and their neighborhood in the context
of PPIs, our model's negative predictions may still have relevance to the immune
system and serve as immune-escape-related proteins under specific conditions.
Despite these limitations, our primary objective remains to illuminate the
unknown immune-escape protein space. Through the utilization of innovative
machine learning approaches, our models are well-positioned to identify novel
immune-escape proteins that may have been missed by traditional experimental
methods.

Overall, our findings contribute to the growing body of literature on
immune-escape proteins and provide new insights into potential targets for
immunotherapy in lung adenocarcinoma.

\section{Conclusion and Future Work}

This study mainly investigates the potential of multimodal representation
learning for low-data protein function prediction. Multimodal representation
learning has become a major research direction in machine learning, attracting
significant attention from researchers~\cite{Dhillon2020-vj}. While previous
works have mainly focused on designing fusion strategies to explore cross-modal
interactions, these methods are limited in their ability to explicitly model
complex cross-modal interactions~\cite{Mai2023-ko}. In recent years, various
modern fusion strategies have been developed to overcome this limitation,
including tensor fusion multi-view RNN networks~\cite{Cui2020-ye},
modality-translation approaches~\cite{Wu2021-qc}, cross-modal attention
mechanisms ~\cite{Wang2022-ev}, graph-based fusion
algorithms~\cite{Angelou2019-zl}, and explainable fusion
approaches~\cite{Wang2022-ev,Holzinger2021-hi}. Meanwhile, multi-task multimodal
learning frameworks have been proposed to capture the inherent correlation
between different multimodal tasks and to help the model better understand the
multimodal contents~\cite{Mai2023-ko}.

Furthermore, while autoencoders have been applied for representation learning in
various domains, there are more complex methods such as variational autoencoders
(VAEs) and generative adversarial networks (GANs) that have shown great
potential for generating high-quality representations. These advanced models can
offer improved performance for multimodal representation learning tasks.

While sequence-based models have shown promise, there is still much potential
for improving single modality representation methods, particularly in the
largely unexplored field of protein-protein interaction (PPI) representation
learning for protein function prediction. Graph neural networks (GNNs), such as
Graph Convolutional Networks (GCNs)~\cite{Chen2020-jd}, Graph Attention Networks
(GATs), and Graph Isomorphism Networks (GINs)~\cite{Zhong2022-yg}, have shown
promising results in predicting and utilizing PPIs for biomedical tasks.
Applying these advanced graph-based representation learning methods to low-data
protein function prediction is an area that has not been extensively explored
yet. However, it has the potential to lead to significant improvements in
performance and could be an exciting area of research.

In addition, current biomedical text-based representation methods are not ideal
for handling long texts. For instance, the Longformer~\cite{Beltagy2020-yo} and
Reformer~\cite{Beltagy2020-yo,Kitaev2020-wc} models are designed to handle long
sequences of text effectively by combining global and local context. These
models offer an opportunity to enhance the performance of large language models
trained on biomedical data by incorporating global context awareness.

Overall, this study highlights the need for continued research in multimodal
representation learning methods, particularly in the field of protein function
prediction, and suggests potential avenues for further improvement and
exploration in the field. The use of advanced models, such as GNNs, VAEs, GANs
and other generative model types offers exciting possibilities for improving
performance and generating high-quality representations. This study encourages
researchers to explore these areas and further develop modern fusion strategies
for explicitly modeling complex cross-modal interactions. The development of
multi-task multimodal learning frameworks could also help the model better
understand multimodal contents and lead to further improvements in performance.

\section{Methods}

\subsection{A Summary of the Benchmark Dataset and Evaluation Metrics}

Gene Ontology (GO)  a standardized system for the functional annotation of genes
and their products. It is a comprehensive, hierarchical vocabulary of terms that
describe the molecular functions (MF), biological processes (BP), and cellular
components (CC) associated with genes and gene products across different
organisms~\cite{Gene_Ontology_Consortium2021-rz}. In our recent study, we
developed the PROBE (Protein RepresentatiOn BEnchmark) tool consisting of a
dataset and evaluation method used to assess the performance of alternative
protein representation methods including prediction of GO term
annotations~\cite{Unsal2022-jz}. In the PROBE, the ontology-based PFP benchmark
dataset was created by using human proteins and their GO term annotations. The
data was preprocessed based on predetermined filtering rules to improve the
reliability of the annotations. The GO terms were then grouped according to the
number of annotated proteins and the depth of the term in the GO graph, and a
selection of dissimilar terms was chosen from each group. The final dataset
consists of 125 GO terms (Supplementary Table S8, S9), each with a list of
annotated proteins, and is used to evaluate the performance of various methods
for predicting GO terms. Specifically, we obtained human proteins and their
corresponding GO term annotations from reliable databases, excluded
electronically made annotations to ensure data quality, created individual lists
for each GO term, filtered proteins using UniRef clusters to eliminate bias, and
grouped GO terms based on their number of annotated proteins and specificity
(Supplementary Table S10). By applying a meticulous approach, we established 27
distinct GO term groups and selected five dissimilar terms from each group to
construct a total of 500 prediction models. Additionally, we incorporated three
rule/association-based methods and evaluated performance through 5-fold
cross-validation. Detailed information on the dataset and the evaluation methods
can be found in the next subsections.

In our study, we utilized weighted F1 scores as well as accuracy, hamming
distance, precision and recall to assess the performance of protein
representations in the context of protein function prediction. The weighted F1
score is a measure of the effectiveness of a classifier that takes into account
the relative importance of each label, as well as the number of true positive,
false positive, and false negative predictions made by the classifier. This
measure is particularly useful in a multi-label setting such as GO term
prediction, where multiple GO terms may be assigned to a single protein, and the
relative importance of each term may vary. To calculate the weighted F1 score,
we first determined the F1 score for each GO term. The weighted F1 score is then
calculated as the mean of all per-class F1 scores, with the weights for each
term determined based on the support of the class.

\subsection{Calculation of Sequence-based Representations}

In our study, we generated protein sequence-based representations by adopting
the results of our recent  PROBE (Protein RepresentatiOn BEnchmark)
study~\cite{Unsal2022-jz}. To obtain the protein sequences for our analysis, we
used the UniProt database~\cite{UniProt_Consortium2019-fp}, which is a widely
used resource for gathering standardized protein sequences. We used canonical
protein sequences which allowed us to ensure the consistency of our calculations
across all proteins.

To generate the sequence representations, we employed 24 representation learning
methods, including both classical (e.g., alignment-based, physicochemical) and
artificial learning-based approaches (Supplementary Table S1a). Artificial
learning-based methods (15 of 24) took the protein sequences as input and
utilized techniques such as deep learning and natural language processing to
extract relevant features of the proteins. These features were then mapped onto
a continuous vector space using techniques like word embeddings or neural
network architectures. The resulting feature vectors captured various inherent
properties of the proteins, including structural and functional information.

It is worth noting that some of the representation learning methods we used in
this study were specifically designed to generate protein-level representations,
while others were intended to generate residue-level representations. For
residue-level representations, we utilized the mean pooling technique to combine
the information from all the residues in the protein and obtain overall protein
representations. Additionally, we carefully considered the source data and
algorithmic approaches used by the various representation learning methods in
order to cover a wide range of methodologies and ensure the comprehensiveness of
our results. Details can be found in the methods section of the PROBE study.
Detailed information about sequence representations that we utilized in our
study can be found in the Methods section.

\subsection{Calculation of Protein-Protein Interaction-based Representations}

Graphs are a powerful way to represent information that has connections, and PPI
data is a good example of this. In PPI data, proteins are represented as nodes
and their interactions as edges. Machine learning algorithms can then be applied
to these graph representations to make predictions about protein interactions.
One common way to do this is by transforming the graph into a low-dimensional
vector representation, which allows the algorithm to generalize the input data
and make predictions. These representations, called graph embeddings, can be
created using various techniques such as Node2Vec~\cite{Grover2016-bo},
HOPE~\cite{Ou2016-er} in which we used this study.

Node2Vec learns vector representations in a similar way to how
word2vec~\cite{Mccormick2016-uk} learns the vector representations of words
(i.e. both use the skip-gram model~\cite{Mccormick2016-uk}).

\noindent The Node2Vec method consists of three steps:

\begin{description}
    \item[Sampling:] A graph is sampled using random walks.
    \item[Training skip-gram:] The random walks used in the sampling process can
    be thought of as a sentence made up of words, where each node represents a
    word and each walk results in a sequence of nodes. These node sequences are
    used for the training of the skip-gram model.
    \item[Calculation of representation vectors:] Protein representations are
    the outputs of the hidden layer in the skip-gram model. When a relevant node
    is given as input, the output is a vector that represents that node. In this
    study, each node represents a protein and each edge represents an
    interaction between these proteins.
\end{description}

Node2Vec models each node as a protein, and each sequence of nodes visited
during a walk is modeled as a sequence. Node2Vec creates node sequences using
the Breadth First Search (BFS)and Depth First Search (DFS)~\cite{Grover2016-bo}
search algorithms. The innovative part of Node2Vec is its sampling strategy,
which it performs using a random walk strategy on the graph. Graph traversal is
the act of moving intentionally along the nodes and edges of a graph for a
specific purpose, while preserving the neighborhoods and structures. To do this,
an appropriate path is needed to explore the graph. BFS and DFS use second-order
random walks to discover neighbors, discover distant neighborhoods and sample
the graph. BFS starts from the starting node and expands outward to all nodes at
the same distance before moving on to the next. DFS starts from the starting
node and explores as far as possible along each branch before backtracking.
Node2Vec uses a biased random walk strategy (tuned via hyperparameters) that
allows the search to focus on the structure of the graph. Node2Vec has two
hyperparameters, $p$ and $q$, that control the sampling strategy. $p$ controls
the return probability (i.e. probability of returning to the previous node) and
$q$ controls the probability of moving to a neighbor node. If $p < q$, the
search is more likely to continue moving outward from the current node, and if
$p > q$, it is more likely to search around the local area. When $p = q$, the
search is unbiased and has an equal chance of moving outward or returning. In
other terms these parameters determine exploration of local and global context.

In the second step Node2Vec uses the skip-gram model to generate representation
vectors. The model has a well known general NN structure. Firstly, the model is
fed forward with the input data. The loss is calculated, and then the weights
are updated based on this loss. This process is repeated for a certain number of
epochs. In this process, $w(t)$ is the input, or target word. There is a hidden
layer where a projection takes place between the input vector $w(t)$ and the
weight matrix. The output of this operation in the hidden layer is then
transferred to the output layer. The output layer performs the calculation
between the output vector of the hidden layer and the weight matrix of the
output layer. Then, in the given context, the softmax activation function is
applied to calculate the probability of the words within the relevant window for
the input $w(t)$. As a result, when any node is taken as input, a vector of size
equal to the number of unique nodes containing the probability values of the
nodes is obtained. Additionally, $\mathbf{V}$ is the known as the node
distribution of the system in the node sequence, and $N$ is the number of
neurons in the hidden layer. The size of an input vector is $|\mathbf{V}|$, and
each node is encoded using one-hot encoding. The weight matrix for the hidden
layer ($\mathbf{W}$) is of size $|\mathbf{V}| \times N$. The product of
$\mathbf{W}$ and the hidden layer gives a output vector of size $|\mathbf{V}|$.

The HOPE (Higher-Order Proximity Embedding) method is a graph representation
technique that aims to preserve the second-order proximity between nodes, which
is the relationship between the neighbors of a node. The HOPE method preserves
this proximity by considering the sum of the singular values of the submatrices
formed by the rows and columns of the adjacency matrix corresponding to the
neighbors of a node. To obtain the vector representations, the HOPE method
reduces the resulting matrix to a $d$-dimensional vector using singular value
decomposition (SVD). This allows the method to capture the inherent structure of
the graph in a low-dimensional space, making it suitable for various downstream
tasks such as node classification and link prediction.

One advantage of the HOPE method is that it can be applied to both undirected
and directed graphs, making it a versatile tool for representing a wide range of
complex data structures. Additionally, the method is computationally efficient,
making it suitable for large-scale graph representation learning tasks. Existing
methods are unable to preserve the critical feature of directed graphs, known as
asymmetric transitivity. Asymmetric transitivity refers to the correlation
between directed edges, meaning that if there is a directed path from $u$ to
$v$, then there is likely to be a directed edge from $u$ to $v$. To overcome
this challenge, the preservation of asymmetric transitivity is based on the use
of High-Order Proximity preserved Embedding (HOPE). The hyperparameter $\beta$
in HOPE is a distortion parameter. It determines how quickly the weight of a
path decreases as the path length increases. 

In the present study, the HOPE algorithm was implemented in Python using the
NetworkX~\cite{Hagberg2008-ub} library. The resulting vector representations
were used as input for prediction models in the study. Overall, the HoPE method
provided robust and informative vector representations that captured the
inherent structure of the protein-protein interaction graphs.

The Node2Vec algorithm was applied to protein-protein interaction graphs
obtained from the IntACT database\footnote{The {\ttfamily intact.zip} file was
downloaded on 07.07.2020 from the IntAct Molecular Interaction Database.
Preprocessing was performed on the {\ttfamily intact.txt} file within the zip
file.} The original size of the file was 1063382 rows and 42 columns. The ID(s)
interactor A and ID(s) interactor B columns were filtered, reducing the data to
two dimensions (1,063,382 rows by 2 columns).

If the both interactions were not represented by uniprot ids, these lines were
filtered. Also, duplicates were then removed, and the remainder was 587,777
interactions. Moreover, we only select human proteins which left 16,435 nodes
(proteins) and 241,833 interactions.

We used our benchmarking tool PROBE to assess alternative hyper parameter
combinations of HOPE and Node2Vec. The hyper parameters used are shown in
Supplementary Table S5 and Supplementary Table S6 in the supplementary material.
We calculated alternative representations for the same protein dataset utilizing
each hyper parameter tuned model. Afterwards, we calculated F1 scores for each
alternative representation. All combinations of dimension ($d$) and $\beta$ were
tested for HOPE and all combinations of $d$, $p$ and $q$ were tested for
Node2Vec. We present results of the best 10 representations of HOPE and Node2Vec
in Figure \ref{fig:PPI_perf} and Supplementary Table S1b.

\subsection{Calculation of Text Representations}

Protein-related information is widely available in the literature in the form of
text. However, the systematic processing of this data remains an open problem.
In our study, we aimed to process this data using the UniProtKB/Swiss-Prot
database and PubMed as data sources. The UniProtKB/Swiss-Prot database is
particularly useful because the information it contains has been summarized by
experts from the literature and entered into the UniProtKB/Swiss-Prot database.

As part of our study, we downloaded all protein records (entries) for all
species in xml format from the UniProtKB/Swiss-Prot database as of June 2020.
These species include archaea, bacteria, fungi, humans, invertebrates, mammals,
plants, rodents, vertebrates, viruses, and unclassified species. A total of
562,253 records were downloaded. For each of these records, we created text
files with the file headers being the UniProt IDs.

We then extracted the ``Function'', ``Cofactor'', ``Subunit'', ``Tissue
Specificity'', ``Induction'', ``Domain'', ``PTM'', and ``Disease'' sub-sections
of the ``Comments'' section (i.e., {\ttfamily <comment>}) of the downloaded
records using the BioPython library. These subsections were saved in text format
one after the other in separate files for each record. Of the downloaded
records, 20,365 were identified as belonging to humans and we use these records
to generate the text based protein representation vectors. We then used the
UniProt IDs of these human protein records to prepare the text information for
creating protein representations from the previously created folder containing
the text information for all records. We also access PubMed papers referenced in
these fields and download their abstracts. These abstracts were saved under a
single file belonging to proteins. We created two datasets, \emph{(i)} text
gathered from UniProt, \emph{(ii)} text gathered from UniProt and PubMed. Hence
the vectors calculated using these two datasets were named as
UNIPROT-VectorModelName and UNIPROT-PubMed-VectorModelName.

We used the BioBERT~\cite{Lee2020-fu}, BioWordVec~\cite{Zhang2019-ya}, and
BioSentVec~\cite{Chen2019-qo} models to process the text information and create
protein representation vectors. BioBERT is a pre-trained language model based on
the BERT model, which was trained on a large biomedical corpus and fine-tuned on
various biomedical tasks. BioWordVec and BioSentVec are word and sentence
embedding models, respectively, that were trained on a large biomedical corpus.
We used the BioBERT model to obtain contextualized word embeddings for each word
in the text. The BioWordVec model was used to obtain word embeddings, and the
BioSentVec model was used to obtain sentence embeddings. These embeddings were
then summerised to form the final protein representation vectors. For this
summarisation we used mean pooling over word vectors.

BioWordVec is a biomedical word embedding method that combines subword
information from unlabeled biomedical text with a widely-used biomedical
controlled vocabulary called Medical Subject Headings (MeSH). It is based on the
fastText model and was trained on PubMed and MIMIC-III Clinical Database text
corpora. The dimensions of the resulting word embeddings are $200$, and the
model was trained with a window size of $20$, a learning rate of $0.05$, a
sampling threshold of $10^{-4}$, and negative examples set to 10. BioWordVec has
been shown to improve performance over the previous state of the art on a range
of biomedical natural language processing (BioNLP) tasks.

BioSentVec is a biomedical sentence embedding method that is based on the
sent2vec model. It was trained on the PubMed and MIMIC-III Clinical Database
text corpora and produces 700-dimensional sentence embeddings. The model uses
the bi-gram model and was trained with a window size of 20 and negative examples
set to 10. BioSentVec has been shown to outperform other unsupervised and
supervised methods on clinical sentence pair similarity tasks using the BIOSSES
and MedSTS datasets. It has also been used to improve performance on a range of
BioNLP tasks.

The BioWordVec and BioSentVec model was trained on the PubMed and MIMIC-III
Clinical Database text corpora, with a total of 28,714,373 documents,
181,634,210 sentences, and 4,354,171,148 tokens in the PubMed corpus, and
2,083,180 documents, 41,674,775 sentences, and 539,006,967 tokens in the
MIMIC-III Clinical notes corpus

Both BioWordVec and BioSentVec are useful for a variety of tasks in biomedical
natural language processing, including document classification, information
retrieval, and text summarization. 

We used term frequency-inverse document frequency (TF-IDF) with principal
component analysis (PCA) to generate representation vectors with dimensions of
256, 512, 768, and 1024. TF-IDF is a statistical measure that reflects the
importance of a word in a document within a corpus, taking into account the
frequency of the word in the document and the inverse frequency of the word
across the entire corpus. By weighting the words in this way, TF-IDF can be used
to identify the most important words in a document and filter out less important
words. PCA is a statistical technique that is used to reduce the dimensionality
of data by identifying the underlying patterns in the data and projecting the
data onto a lower-dimensional space. In our study, we used PCA to reduce the
dimensionality of the representation vectors generated from the TF-IDF matrix.
By reducing the dimensionality of the vectors, we were able to capture the most
important features of the data while eliminating less important features.

\subsection{Calculation of Multimodal Representations}

In this study, we develop three autoencoder-based methods for combining protein
representations. Here, the input representation vectors are produced using text,
aminoacid sequences, and protein-protein interactions which are denoted as
$\mathbf{i}_t$, $\mathbf{i}_s$, $\mathbf{i}_p$ respectively. Similarly, output
representation vectors are denoted as $\mathbf{o}_t$, $\mathbf{o}_s$,
$\mathbf{o}_p$.

The first method is a simple autoencoder (SimpleAE) that takes the concatenated
form of the individual representation vectors as input. The middle layer of the
trained autoencoder is intended to be used as a vector representing the protein,
and is denoted as $\mathbf{z}_{mid}$. The input ($\mathbf{v}_i$) and output
($\mathbf{v}_o$) vectors in this model were concatenated as $\mathbf{v}_i =
(\mathbf{i}_t, \mathbf{i}_s, \mathbf{i}_p)$ and $\mathbf{v}_o =  (\mathbf{o}_t,
\mathbf{o}_s, \mathbf{o}_p)$.





The second method is a multimodal autoencoder (MultiModalAE) that involves the
reconstruction of individual representation vectors for the sequence, text, and
protein-protein interaction data separately. In this case, again,
$\mathbf{z}_{mid}$ is also used as a vector representing the protein. Here the
difference is each modality that each input is propagated independently up to
the $\mathbf{z}_{mid}$ layer. After the $\mathbf{z}_{mid}$ layer, the modalities
are again separated (see Figure~\ref{fig:multimodal}b ). 



$\mathbf{z}_{mid}$ was calculated by integrating each modality in a fully connected manner;

\begin{equation}
    \mathbf{z}_{mid} = f_{text} \odot  f_{sequence} \odot f_{PPI} 
\end{equation}

The third method, called sequence-based autoencoder (TransferAE), is created
using transfer learning, in which the weights of a trained MultiModalAE model
are transferred to another autoencoder model that takes protein sequence
representation as input and generates representations for the text and
protein-protein interaction data. This method allows for the generation of
protein representations for proteins for which data is only available in the
form of sequence information.

The TransferAE model is created using transfer learning from the MultiModalAE
model. Since the TransferAE model takes only sequence data as input, it starts
as a subnetwork of the MultiModalAE model, excluding input layers for PPI, and
text. Then the TransferAE is then further trained with sequence input only, so
that it can learn to reconstruct the representation vectors for text and PPIs
using only this input. The input sequence is also output by the network, for
verification and error backpropagation. Like the MultiModalAE model, the
TransferAE model also has multiple hidden layers with varying dimensions. For
this model we transfer weights and biases of sequence input.


However for the decoder the output vectors are still $\mathbf{o}_t$,
$\mathbf{o}_s$, $\mathbf{o}_p$. Here, we should note that decoder weights and
biases for these modalities were transferred from MultiModalAE to TransferAE.

Alternative $\mathbf{z}_{mid}$ dimensions (256, 512, 768, 1024) were tested for
all models, in order to determine the optimal size for the representation
vectors. The models are trained using the text sequence, PPI, and protein
sequence representation vectors, and the performance of the models is evaluated
using a weighted F1 score. 

For all AE models we used the ReduceLROnPlateau function from the lr\_scheduler
module of PyTorch~\cite{Imambi2021-ym}. This function allows us to reduce the
learning rate of the optimizer when the performance of the model plateaus, or
stops improving. This can help prevent overfitting, and allows the model to
continue learning and improving even when it is no longer making progress with
the current learning rate.

The optimizer used in these models is the AdamW optimizer, which is a variant of
the popular Adam optimizer that includes weight decay, which can help prevent
overfitting by regularizing the model's weights. The learning rate for the AdamW
optimizer is set to 1e-3, which is a common choice for many deep learning
models.

The loss function used in these models is the mean squared error (MSE)
loss~\cite{Harville1992-vj}, which is a common choice for regression tasks. MSE
loss is calculated by taking the squared difference between the predicted and
true values, and averaging the squared differences over all examples in the
dataset. This loss function is sensitive to large errors, and can be effective
at encouraging the model to make small, precise predictions.

Overall, the combination of the ReduceLROnPlateau function, AdamW optimizer, and
MSE loss function can be effective at training deep learning models for
regression tasks, such as protein representation generation. These components
can help the model learn effectively, prevent overfitting, and make accurate
predictions.

\subsection{The Methodology for the Immune-escape Protein Prediction}

This case study aimed to predict new immune escape proteins in lung
adenocarcinomas, a cancer type with a high mortality rate, using a prediction
model with a low amount of training data. The study utilized TCGA-LUAD RNA-Seq
data to calculate differentially expressed genes, resulting in 1434 genes, of
which 329 were low, and 1105 were highly expressed in lung adenocarcinoma tumors
compared to the matched normal lung tissue (Supplementary File 1). A negative
dataset was created using GO BP ontology-based similarities between proteins and
semantic similarity calculations to known immune-escape proteins. Multimodal
protein representation vectors with autoencoders were generated using protein
sequence, PPI, and text data. Finally, the study utilized autoencoder-based
representation to train immune-escape predictors models using fully connected
feed-forward neural networks (FC-NN) to predict new immune-escape proteins. The
study predicted 121 immune-escape proteins (Supplementary Table S3) from the
differentially expressed gene set, and the predictions were evaluated via
literature search and existing functional annotations in source databases. 

The aim of the first phase was to identify differentially expressed genes
between tumor and normal tissue samples for the purpose of selecting
immune-escape protein candidates. To achieve this goal, the TCGA-LUAD (The
Cancer Genome Atlas Lung Adenocarcinoma) data was downloaded from the TCGA
database~\cite{Cancer_Genome_Atlas_Research_Network2014-oy} using the
TCGABiolinks package~\cite{Mounir2019-tz}. The RNA-Seq method, with Illumina
HiSeq as the sequencing platform, was chosen as the experimental method for the
data set. TCGABiolinks allows for searching, downloading, and analyzing data
from the National Cancer Institute database using the GDC Application
Programming Interface (API) with R programming~\cite{Gentleman2004-fg} .

Here we first filtered data based on read count and then modeled data using the
General Linear Model (GLM). The "quantile" method was used in the filtering
phase. To use this method, a cut-off value for the read count in RNA-Seq
analysis must be determined. In this study, this value was set to 0.25
(qnt.cut=0.25). Thus, the 25th percentile of the data frame representing each
row as a gene was determined. The "glmLRT" method was used in the differential
gene expression analysis phase~\cite{Chen_undated-iz}. This model fits the read
counts to a negative binomial distribution~\cite{Ren2020-qh}  and performs
ANOVA-style tests~\cite{Girden1992-mo}. The "glmLRT" method used in this study,
using the GLM, checks for differences between groups to determine whether there
is a difference in expression between tumor and normal tissue. After the data
distribution model is created with GLM in this phase, the hypothesis test is
performed with the log-odds ratio. The p-value and adjusted p-value (false
discovery rate -- FDR) for each gene are calculated in this test. 

The immune-escape protein candidates were selected using gene expression
analysis with the aim of identifying tumor immune-escape proteins among the
proteins encoded by differentially expressed genes. The TCGA-LUAD data was
downloaded from the TCGA database using the TCGABiolinks package and the RNA-Seq
method with the Illumina HiSeq platform was chosen as the experimental method
for the data set. The "quantile" and "glmLRT'' methods were used in the
implementation of gene expression analysis in R. The "quantile" method was used
in the filtering phase, and the "glmLRT'' method was used in the differential
gene expression analysis phase. The genes over 2 fold-change and FDR is smaller
than 0.01 were considered as differentially expressed. This set was used as
input for the immune-escape prediction model later.

The positive samples in the training set were gathered by an exhaustive
literature search. Which are; IL10, IL2, CD44, TGFB1, IL12A, SELL, PTGES2, IFNG,
EBAG9, VEGFA, CXCR1, CXCR2, LMP1, CTLA4, ARG1, PSMB8, IDO1, HLA-G, TAP1, ICAM1,
FASLG, IKBKB, CD80, CD274, CD44, CAMP, TNFRSF6B, FOXP3, CD47~\cite{Brandau2011,
Jadus2012, Hunn2016, Vinay2015, Leone2018, Huang2017, Carosella2015,
Zelenay2015, Kotteas2014})

A set of negative examples was created for the training of a classifier. This
set was obtained from To this end, a semantic similarity matrix based on the
Gene Ontology (GO) was first constructed. The Lin similarity~\cite{Lin1998-je}
from the GoSemSim package~\cite{Yu2010-nc} was used to independently calculate
the real GO-based semantic similarities between all proteins in the BP
(biological process) category. Only protein-GO tags approved by human curators
were used in these similarity calculations. The Lin similarity is based on
Shannon's information theory, which states that the information content (IC) of
an event is inversely proportional to the probability (P) of observing the
event.  Another concept used in the Lin similarity is the least common subsumer
(LCS), which is the first common ancestor of two GO terms in the hierarchical GO
graph when going to the root.  The BP-based similarity matrix contains 6154
proteins. There are 26 immune-escape proteins (Supplementary Table S11) that can
be represented by the vector of any of the three protein representation methods
in the immune-escape list. First, the proteins for which the vector was
calculated by all three protein representation methods were selected from the
6154 proteins. Subsequently, these proteins were filtered to ensure that they
had a maximum of 50\% sequence similarity with UniRef50~\cite{Suzek2015}.
After this process, half of the remaining proteins with the highest average
semantic distance from the immune-escape proteins were selected, and 124
proteins were randomly chosen from this group (Supplementary Table S12). In this
way, a set of 124 proteins that are assumed not to participate in immune-escape
processes was obtained. Overall, there are 26 positive and 124 negative samples
used for training with available text, sequence and PPI data.

On the immune-escape prediction model training we first create a baseline model.
We utilized SeqVec for protein sequence representation, BioBERT for text
representation, and Node2Vec for protein-protein interaction (PPI)
representation. We have selected our best autoencoder-based representation
(i.e., MultiModalAE-256) for low-data on the GO BP category. We utilized fully
connected feed-forward neural networks (FC-NN). For the model we calculate,
precision, recall, F1-Micro, F1-Macro, F1-Weighted and F-Max scores. We trained
the model within a hyperparameter search using 5-Fold cross validation. During
the hyperparameter search the goal was maximizing the F-Max score. At the end of
the parameter search we refit each model with all data before predicting the
immune-escape protein candidates. The detailed results can be found in the
Supplementary Table S7.

\printbibliography

@article{Schwarzer2020-cg,
  title    = {{Data-Efficient} Reinforcement Learning with {Self-Predictive}
              Representations},
  author   = {Schwarzer, Max and Anand, Ankesh and Goel, Rishab and Hjelm, R
              Devon and Courville, Aaron and Bachman, Philip},
  journal  = {arXiv preprint arXiv:2007.05929},
  month    = jul,
  year     = 2020
}

@article{Bepler2021-yc,
  title     = {Learning the protein language: Evolution, structure, and
               function},
  author    = {Bepler, Tristan and Berger, Bonnie},
  journal   = {Cell Systems},
  publisher = {Cell Press},
  volume    = 12,
  number    = 6,
  pages     = {654--669.e3},
  month     = jun,
  year      = 2021
}

@article{Zhang2019-ya,
  title    = {{BioWordVec}, improving biomedical word embeddings with subword
              information and {MeSH}},
  author   = {Zhang, Yijia and Chen, Qingyu and Yang, Zhihao and Lin, Hongfei
              and Lu, Zhiyong},
  journal  = {Sci Data},
  volume   = 6,
  number   = 1,
  pages    = {52},
  month    = may,
  year     = 2019,
  language = {en}
}

@book{Girden1992-mo,
  title     = {{ANOVA}: Repeated Measures},
  author    = {Girden, Ellen R},
  publisher = {SAGE},
  year      = 1992,
  language  = {en}
}

@techreport{Hagberg2008-ub,
  title       = {Exploring network structure, dynamics, and function using
                 networkx},
  author      = {Hagberg, Aric and Swart, Pieter and S Chult, Daniel},
  publisher   = {osti.gov},
  number      = {LA-UR-08-05495; LA-UR-08-5495},
  institution = {Los Alamos National Lab. (LANL), Los Alamos, NM (United
                 States)},
  month       = jan,
  year        = 2008,
  language    = {en}
}

@article{Li2021-yd,
  title    = {{Capsule-LPI}: a {LncRNA-protein} interaction predicting tool
              based on a capsule network},
  author   = {Li, Ying and Sun, Hang and Feng, Shiyao and Zhang, Qi and Han,
              Siyu and Du, Wei},
  journal  = {BMC Bioinformatics},
  volume   = 22,
  number   = 1,
  pages    = {246},
  month    = may,
  year     = 2021,
  language = {en}
}

@article{Scheunemann2019-id,
  title         = {Effect of Imbalanced Charge Transport on the Interplay of
                   Surface and Bulk Recombination in Organic Solar Cells},
  author        = {Scheunemann, Dorothea and Wilken, Sebastian and Sandberg,
                   Oskar J and {\"O}sterbacka, Ronald and Schiek, Manuela},
  month         = may,
  year          = 2019,
  archiveprefix = {arXiv},
  primaryclass  = {physics.app-ph},
  eprint        = {1905.01268}
}

@article{Liang2021-rx,
  title         = {{MultiBench}: Multiscale Benchmarks for Multimodal
                   Representation Learning},
  author        = {Liang, Paul Pu and Lyu, Yiwei and Fan, Xiang and Wu, Zetian
                   and Cheng, Yun and Wu, Jason and Chen, Leslie and Wu, Peter
                   and Lee, Michelle A and Zhu, Yuke and Salakhutdinov, Ruslan
                   and Morency, Louis-Philippe},
  month         = jul,
  year          = 2021,
  archiveprefix = {arXiv},
  primaryclass  = {cs.LG},
  eprint        = {2107.07502}
}

@inproceedings{Vig2022-kr,
  title     = {{BERTology} Meets Biology: Interpreting Attention in Protein
               Language Models},
  booktitle = {International Conference on Learning Representations},
  author    = {Vig, Jesse and Madani, Ali and Varshney, Lav R and Xiong,
               Caiming and Socher, Richard and Rajani, Nazneen},
  month     = feb,
  year      = 2022
}

@article{Meier2021-ny,
  title   = {Language models enable zero-shot prediction of the effects of
             mutations on protein function},
  author  = {Meier, Joshua and Rao, Roshan and Verkuil, Robert and Liu, Jason
             and Sercu, Tom and Rives, Alex},
  journal = {Adv. Neural Inf. Process. Syst.},
  volume  = 34,
  month   = dec,
  year    = 2021
}

@article{Ashburner2000-fo,
  title     = {Gene Ontology: tool for the unification of biology},
  author    = {Ashburner, Michael and Ball, Catherine A and Blake, Judith A and
               Botstein, David and Butler, Heather and Cherry, J Michael and
               Davis, Allan P and Dolinski, Kara and Dwight, Selina S and
               Eppig, Janan T and Harris, Midori A and Hill, David P and
               Issel-Tarver, Laurie and Kasarskis, Andrew and Lewis, Suzanna
               and Matese, John C and Richardson, Joel E and Ringwald, Martin
               and Rubin, Gerald M and Sherlock, Gavin},
  journal   = {Nat. Genet.},
  publisher = {Nature Publishing Group},
  volume    = 25,
  number    = 1,
  pages     = {25--29},
  month     = may,
  year      = 2000,
  language  = {en}
}

@article{Mounir2019-tz,
  title    = {New functionalities in the {TCGAbiolinks} package for the study
              and integration of cancer data from {GDC} and {GTEx}},
  author   = {Mounir, Mohamed and Lucchetta, Marta and Silva, Tiago C and
              Olsen, Catharina and Bontempi, Gianluca and Chen, Xi and
              Noushmehr, Houtan and Colaprico, Antonio and Papaleo, Elena},
  journal  = {PLoS Comput. Biol.},
  volume   = 15,
  number   = 3,
  pages    = {e1006701},
  month    = mar,
  year     = 2019,
  language = {en}
}

@article{Beltagy2020-yo,
  title         = {Longformer: The {Long-Document} Transformer},
  author        = {Beltagy, Iz and Peters, Matthew E and Cohan, Arman},
  month         = apr,
  year          = 2020,
  archiveprefix = {arXiv},
  primaryclass  = {cs.CL},
  eprint        = {2004.05150}
}

@article{Makrodimitris2020-vb,
  title     = {Automatic Gene Function Prediction in the 2020's},
  author    = {Makrodimitris, Stavros and van Ham, Roeland C H J and Reinders,
               Marcel J T},
  journal   = {Genes},
  publisher = {Multidisciplinary Digital Publishing Institute},
  volume    = 11,
  number    = 11,
  pages     = {1264},
  month     = oct,
  year      = 2020,
  language  = {en}
}

@article{Wu2021-qc,
  title    = {{BABEL} enables cross-modality translation between multiomic
              profiles at single-cell resolution},
  author   = {Wu, Kevin E and Yost, Kathryn E and Chang, Howard Y and Zou,
              James},
  journal  = {Proceedings of the National Academy of Sciences},
  volume   = 118,
  number   = 15,
  pages    = {e2023070118},
  year     = 2021
}

@inproceedings{Tuggener2020-zl,
  title     = {{LEDGAR}: A {Large-Scale} Multi-label Corpus for Text
               Classification of Legal Provisions in Contracts},
  booktitle = {Proceedings of the Twelfth Language Resources and Evaluation
               Conference},
  author    = {Tuggener, Don and Von D{\"a}niken, Pius and Peetz, Thomas and
               Cieliebak, Mark},
  pages     = {1235--1241},
  year      = 2020
}

@inproceedings{Stanley2021-mn,
  title     = {{FS-Mol}: A {Few-Shot} Learning Dataset of Molecules},
  booktitle = {Thirty-fifth Conference on Neural Information Processing Systems
               Datasets and Benchmarks Track (Round 2)},
  author    = {Stanley, Megan and Bronskill, John F and Maziarz, Krzysztof and
               Misztela, Hubert and Lanini, Jessica and Segler, Marwin and
               Schneider, Nadine and Brockschmidt, Marc},
  month     = nov,
  year      = 2021
}

@article{Zheng2021-gu,
  title    = {Deep scaffold hopping with multimodal transformer neural networks},
  author   = {Zheng, Shuangjia and Lei, Zengrong and Ai, Haitao and Chen,
              Hongming and Deng, Daiguo and Yang, Yuedong},
  journal  = {J. Cheminform.},
  volume   = 13,
  number   = 1,
  pages    = {87},
  month    = nov,
  year     = 2021,
  language = {en}
}

@article{Zhang2019-un,
  title    = {Multimodal deep representation learning for protein interaction
              identification and protein family classification},
  author   = {Zhang, Da and Kabuka, Mansur},
  journal  = {BMC Bioinformatics},
  volume   = 20,
  number   = {Suppl 16},
  pages    = {531},
  month    = dec,
  year     = 2019,
  language = {en}
}

@article{Lee2020-fu,
  title    = {{BioBERT}: a pre-trained biomedical language representation model
              for biomedical text mining},
  author   = {Lee, Jinhyuk and Yoon, Wonjin and Kim, Sungdong and Kim,
              Donghyeon and Kim, Sunkyu and So, Chan Ho and Kang, Jaewoo},
  journal  = {Bioinformatics},
  volume   = 36,
  number   = 4,
  pages    = {1234--1240},
  month    = feb,
  year     = 2020,
  language = {en}
}

@inproceedings{Lin1998-je,
  title     = {An information-theoretic definition of similarity},
  booktitle = {Icml},
  author    = {Lin, Dekang and {Others}},
  volume    = 98,
  pages     = {296--304},
  year      = 1998
}

@article{Gligorijevic2018-kh,
  title    = {{deepNF}: deep network fusion for protein function prediction},
  author   = {Gligorijevic, Vladimir and Barot, Meet and Bonneau, Richard},
  journal  = {Bioinformatics},
  volume   = 34,
  number   = 22,
  pages    = {3873--3881},
  month    = nov,
  year     = 2018,
  language = {en}
}

@article{UniProt_Consortium2019-fp,
  title     = {{UniProt}: a worldwide hub of protein knowledge},
  author    = {{UniProt Consortium}},
  journal   = {Nucleic Acids Res.},
  publisher = {academic.oup.com},
  volume    = 47,
  number    = {D1},
  pages     = {D506--D515},
  month     = jan,
  year      = 2019,
  language  = {en}
}

@article{Holzinger2021-hi,
  title     = {Towards multi-modal causability with Graph Neural Networks
               enabling information fusion for explainable {AI}},
  author    = {Holzinger, Andreas and Malle, Bernd and Saranti, Anna and
               Pfeifer, Bastian},
  journal   = {Inf. Fusion},
  publisher = {Elsevier},
  volume    = 71,
  pages     = {28--37},
  month     = jul,
  year      = 2021
}

@article{Mostavi2021-mx,
  title     = {{CancerSiamese}: one-shot learning for predicting primary and
               metastatic tumor types unseen during model training},
  author    = {Mostavi, Milad and Chiu, Yu-Chiao and Chen, Yidong and Huang,
               Yufei},
  journal   = {BMC Bioinformatics},
  publisher = {BioMed Central},
  volume    = 22,
  number    = 1,
  pages     = {1--17},
  month     = may,
  year      = 2021,
  language  = {en}
}

@article{Van_der_Maaten2008-cu,
  title   = {Visualizing Data using {t-SNE}},
  author  = {van der Maaten, Laurens and Hinton, Geoffrey},
  journal = {J. Mach. Learn. Res.},
  volume  = 9,
  number  = 86,
  pages   = {2579--2605},
  year    = 2008
}

@article{Wang2019-fs,
  title         = {What Makes Training {Multi-Modal} Classification Networks
                   Hard?},
  author        = {Wang, Weiyao and Tran, Du and Feiszli, Matt},
  month         = may,
  year          = 2019,
  archiveprefix = {arXiv},
  primaryclass  = {cs.CV},
  eprint        = {1905.12681}
}

@misc{Friedberg2006-oh,
  title   = {Automated protein function prediction--the genomic challenge},
  author  = {Friedberg, I},
  journal = {Briefings in Bioinformatics},
  volume  = 7,
  number  = 3,
  pages   = {225--242},
  year    = 2006
}

@article{Xu2021-rc,
  title         = {{MUFASA}: Multimodal Fusion Architecture Search for
                   Electronic Health Records},
  author        = {Xu, Zhen and So, David R and Dai, Andrew M},
  month         = feb,
  year          = 2021,
  archiveprefix = {arXiv},
  primaryclass  = {cs.LG},
  eprint        = {2102.02340}
}

@misc{Elnaggar2022-ja,
  title        = {{ProtTrans}: Towards Cracking the Language of Lifes Code
                  Through {Self-Supervised} Deep Learning and High Performance
                  Computing},
  author       = {Elnaggar, Ahmed and Heinzinger, Michael and Dallago,
                  Christian and Rihawi, Ghalia and Wang, Yu and Jones, Llion
                  and Gibbs, Tom and Feher, Tamas and Angerer, Christoph and
                  Steinegger, Martin and Bhowmik, Debsindhu and Rost, Burkhard},
  month        = jul,
  year         = 2022,
  howpublished = {\url{https://ieeexplore.ieee.org/document/9477085}},
  note         = {Accessed: 2022-9-12},
  language     = {en}
}

@article{Angelou2019-zl,
  title     = {Graph-based multimodal fusion with metric learning for
               multimodal classification},
  author    = {Angelou, Michalis and Solachidis, Vassilis and Vretos, Nicholas
               and Daras, Petros},
  journal   = {Pattern Recognit.},
  publisher = {Elsevier},
  volume    = 95,
  pages     = {296--307},
  month     = nov,
  year      = 2019
}

@article{Chen2022-ua,
  title     = {A comparative study of automated legal text classification using
               random forests and deep learning},
  author    = {Chen, Haihua and Wu, Lei and Chen, Jiangping and Lu, Wei and
               Ding, Junhua},
  journal   = {Inf. Process. Manag.},
  publisher = {Pergamon},
  volume    = 59,
  number    = 2,
  pages     = {102798},
  month     = mar,
  year      = 2022
}

@misc{Shehu2016-dm,
  title   = {A Survey of Computational Methods for Protein Function Prediction},
  author  = {Shehu, Amarda and Barbar{\'a}, Daniel and Molloy, Kevin},
  journal = {Big Data Analytics in Genomics},
  pages   = {225--298},
  year    = 2016
}

@article{Kitaev2020-wc,
  title         = {Reformer: The Efficient Transformer},
  author        = {Kitaev, Nikita and Kaiser, {\L}ukasz and Levskaya, Anselm},
  month         = jan,
  year          = 2020,
  archiveprefix = {arXiv},
  primaryclass  = {cs.LG},
  eprint        = {2001.04451}
}

@article{Ren2020-qh,
  title    = {Negative binomial additive model for {RNA-Seq} data analysis},
  author   = {Ren, Xu and Kuan, Pei-Fen},
  journal  = {BMC Bioinformatics},
  volume   = 21,
  number   = 1,
  pages    = {171},
  month    = may,
  year     = 2020,
  language = {en}
}

@article{Liu2021-yc,
  title    = {Novel {Few-Shot} Learning Neural Network for Predicting
              {Carbohydrate-Active} Enzyme Affinity Toward
              {Fructo-Oligosaccharides}},
  author   = {Liu, Shaoxun and Kou, Yi and Chen, Lin},
  journal  = {J. Comput. Biol.},
  volume   = 28,
  number   = 12,
  pages    = {1208--1218},
  month    = dec,
  year     = 2021,
  language = {en}
}

@article{Zeng2022-vo,
  title    = {Prognostic and Immunological Roles of {MMP-9} in {Pan-Cancer}},
  author   = {Zeng, Yudan and Gao, Mengqian and Lin, Dongtao and Du, Guoxia and
              Cai, Yongming},
  journal  = {Biomed Res. Int.},
  volume   = 2022,
  pages    = {2592962},
  month    = feb,
  year     = 2022,
  language = {en}
}

@article{Deznabi2020-yi,
  title    = {{DeepKinZero}: zero-shot learning for predicting
              kinase-phosphosite associations involving understudied kinases},
  author   = {Deznabi, Iman and Arabaci, Busra and Koyut{\"u}rk, Mehmet and
              Tastan, Oznur},
  journal  = {Bioinformatics},
  volume   = 36,
  number   = 12,
  pages    = {3652--3661},
  month    = jun,
  year     = 2020,
  language = {en}
}

@article{Hu2020-wy,
  title    = {Learning Multimodal Networks From Heterogeneous Data for
              Prediction of {lncRNA-miRNA} Interactions},
  author   = {Hu, Pengwei and Huang, Yu-An and Chan, Keith C C and You,
              Zhu-Hong},
  journal  = {IEEE/ACM Trans. Comput. Biol. Bioinform.},
  volume   = 17,
  number   = 5,
  pages    = {1516--1524},
  month    = sep,
  year     = 2020,
  language = {en}
}

@article{The_Cancer_Genome_Atlas_Research_Network2012-bw,
  title     = {Comprehensive genomic characterization of squamous cell lung
               cancers},
  author    = {{The Cancer Genome Atlas Research Network}},
  journal   = {Nature},
  publisher = {Nature Publishing Group},
  volume    = 489,
  number    = 7417,
  pages     = {519--525},
  month     = sep,
  year      = 2012,
  language  = {en}
}

@article{Unsal2022-jz,
  title     = {Learning functional properties of proteins with language models},
  author    = {Unsal, Serbulent and Atas, Heval and Albayrak, Muammer and
               Turhan, Kemal and Acar, Aybar C and Do{\u g}an, Tunca},
  journal   = {Nature Machine Intelligence},
  publisher = {Nature Publishing Group},
  volume    = 4,
  number    = 3,
  pages     = {227--245},
  month     = mar,
  year      = 2022,
  language  = {en}
}

@article{Sanh2019-wq,
  title         = {{DistilBERT}, a distilled version of {BERT}: smaller,
                   faster, cheaper and lighter},
  author        = {Sanh, Victor and Debut, Lysandre and Chaumond, Julien and
                   Wolf, Thomas},
  month         = oct,
  year          = 2019,
  archiveprefix = {arXiv},
  primaryclass  = {cs.CL},
  eprint        = {1910.01108}
}

@article{Xu2022-pn,
  title     = {{ProTranslator}: {Zero-Shot} Protein Function Prediction Using
               Textual Description},
  author    = {Xu, Hanwen and Wang, Sheng},
  journal   = {Res. Comput. Mol. Biol.},
  publisher = {Springer International Publishing},
  pages     = {279--294},
  year      = 2022,
  language  = {en}
}

@article{Gat2020-qk,
  title         = {Removing bias in multi-modal classifiers: Regularization by
                   maximizing functional entropies},
  author        = {Gat, Itai and Schwartz, Idan and Schwing, Alexander and
                   Hazan, Tamir},
  pages         = {3197--3208},
  month         = oct,
  year          = 2020,
  copyright     = {http://creativecommons.org/publicdomain/zero/1.0/},
  archiveprefix = {arXiv},
  primaryclass  = {cs.CV},
  eprint        = {2010.10802}
}

@article{Vaswani2017-pi,
  title    = {Attention Is All You Need},
  author   = {Vaswani, Ashish and Shazeer, Noam and Parmar, Niki and Uszkoreit,
              Jakob and Jones, Llion and Gomez, Aidan N and Kaiser, Lukasz and
              Polosukhin, Illia},
  month    = jun,
  year     = 2017
}

@misc{Aizawa2003-hk,
  title   = {An information-theoretic perspective of tf--idf measures},
  author  = {Aizawa, Akiko},
  journal = {Information Processing \& Management},
  volume  = 39,
  number  = 1,
  pages   = {45--65},
  year    = 2003
}

@article{Gene_Ontology_Consortium2021-rz,
  title    = {The Gene Ontology resource: enriching a {GOld} mine},
  author   = {{Gene Ontology Consortium}},
  journal  = {Nucleic Acids Res.},
  volume   = 49,
  number   = {D1},
  pages    = {D325--D334},
  month    = jan,
  year     = 2021,
  language = {en}
}

@article{Khwaja2022-rx,
  title    = {{CELL-E}: Biological {Zero-Shot} {Text-to-Image} Synthesis for
              Protein Localization Prediction},
  author   = {Khwaja, Emaad and Song, Yun S and Huang, Bo},
  journal  = {bioRxiv},
  pages    = {2022.05.27.493774},
  month    = may,
  year     = 2022,
  language = {en}
}

@article{Guo2019-sf,
  title    = {Deep Multimodal Representation Learning: A Survey},
  author   = {Guo, Wenzhong and Wang, Jianwen and Wang, Shiping},
  journal  = {IEEE Access},
  volume   = 7,
  pages    = {63373--63394},
  year     = 2019
}

@misc{Mccormick2016-uk,
  title        = {{Word2Vec} tutorial -the skip-gram model},
  author       = {Mccormick, Chris},
  year         = 2016,
  howpublished = {\url{https://www.fer.unizg.hr/_download/repository/TAR-2020-reading-05.pdf}},
  note         = {Accessed: 2023-3-18}
}

@article{Yao2021-ke,
  title     = {{NetGO} 2.0: improving large-scale protein function prediction
               with massive sequence, text, domain, family and network
               information},
  author    = {Yao, Shuwei and You, Ronghui and Wang, Shaojun and Xiong, Yi and
               Huang, Xiaodi and Zhu, Shanfeng},
  journal   = {Nucleic Acids Res.},
  publisher = {Oxford Academic},
  volume    = 49,
  number    = {W1},
  pages     = {W469--W475},
  month     = may,
  year      = 2021,
  language  = {en}
}

@article{Cancer_Genome_Atlas_Research_Network2014-oy,
  title    = {Comprehensive molecular profiling of lung adenocarcinoma},
  author   = {{Cancer Genome Atlas Research Network}},
  journal  = {Nature},
  volume   = 511,
  number   = 7511,
  pages    = {543--550},
  month    = jul,
  year     = 2014,
  language = {en}
}

@article{Dhillon2020-vj,
  title     = {Systems Biology Approaches to Understanding the Human Immune
               System},
  author    = {Dhillon, Bhavjinder K and Smith, Maren and Baghela, Arjun and
               Lee, Amy H Y and Hancock, Robert E W},
  journal   = {Front. Immunol.},
  publisher = {Frontiers},
  volume    = 11,
  month     = jul,
  year      = 2020,
  language  = {en}
}

@inproceedings{Chen2020-jd,
  title     = {Simple and Deep Graph Convolutional Networks},
  booktitle = {Proceedings of the 37th International Conference on Machine
               Learning},
  author    = {Chen, Ming and Wei, Zhewei and Huang, Zengfeng and Ding, Bolin
               and Li, Yaliang},
  editor    = {Iii, Hal Daum{\'e} and Singh, Aarti},
  publisher = {PMLR},
  volume    = 119,
  pages     = {1725--1735},
  series    = {Proceedings of Machine Learning Research},
  year      = 2020
}

@article{Wilde2019-pz,
  title    = {Immune Dysfunction and {Albumin-Related} Immunity in Liver
              Cirrhosis},
  author   = {Wilde, Benjamin and Katsounas, Antonios},
  journal  = {Mediators Inflamm.},
  volume   = 2019,
  pages    = {7537649},
  month    = feb,
  year     = 2019,
  language = {en}
}

@article{Xue2021-dd,
  title         = {Multimodal {Pre-Training} Model for Sequence-based
                   Prediction of {Protein-Protein} Interaction},
  author        = {Xue, Yang and Liu, Zijing and Fang, Xiaomin and Wang, Fan},
  month         = dec,
  year          = 2021,
  archiveprefix = {arXiv},
  primaryclass  = {q-bio.BM},
  eprint        = {2112.04814}
}

@inproceedings{Tsai2019-np,
  title     = {Multimodal Transformer for Unaligned Multimodal Language
               Sequences},
  booktitle = {Proceedings of the 57th Annual Meeting of the Association for
               Computational Linguistics},
  author    = {Tsai, Yao-Hung Hubert and Bai, Shaojie and Liang, Paul Pu and
               Kolter, J Zico and Morency, Louis-Philippe and Salakhutdinov,
               Ruslan},
  publisher = {Association for Computational Linguistics},
  pages     = {6558--6569},
  month     = jul,
  year      = 2019,
  address   = {Florence, Italy}
}

@article{Mai2023-ko,
  title     = {Excavating multimodal correlation for representation learning},
  author    = {Mai, Sijie and Sun, Ya and Zeng, Ying and Hu, Haifeng},
  journal   = {Inf. Fusion},
  publisher = {Elsevier},
  volume    = 91,
  pages     = {542--555},
  month     = mar,
  year      = 2023
}

@article{Baltrusaitis2019-rv,
  title    = {Multimodal Machine Learning: A Survey and Taxonomy},
  author   = {Baltrusaitis, Tadas and Ahuja, Chaitanya and Morency,
              Louis-Philippe},
  journal  = {IEEE Trans. Pattern Anal. Mach. Intell.},
  volume   = 41,
  number   = 2,
  pages    = {423--443},
  month    = feb,
  year     = 2019,
  language = {en}
}

@article{Li2008-il,
  title    = {{TGF-beta}: a master of all {T} cell trades},
  author   = {Li, Ming O and Flavell, Richard A},
  journal  = {Cell},
  volume   = 134,
  number   = 3,
  pages    = {392--404},
  month    = aug,
  year     = 2008,
  language = {en}
}

@article{kurmi2020-408,
title = {Nitrogen Metabolism in Cancer and Immunity},
journal = {Trends in Cell Biology},
volume = {30},
number = {5},
pages = {408-424},
year = {2020},
issn = {0962-8924},
doi = {https://doi.org/10.1016/j.tcb.2020.02.005},
url = {https://www.sciencedirect.com/science/article/pii/S0962892420300404},
author = {Kiran Kurmi and Marcia C. Haigis}
}

@article{Sureyya_Rifaioglu2019-mw,
  title     = {{DEEPred}: Automated Protein Function Prediction with Multi-task
               Feed-forward Deep Neural Networks},
  author    = {Sureyya Rifaioglu, Ahmet and Do{\u g}an, Tunca and Jesus Martin,
               Maria and Cetin-Atalay, Rengul and Atalay, Volkan},
  journal   = {Sci. Rep.},
  publisher = {Nature Publishing Group},
  volume    = 9,
  number    = 1,
  pages     = {1--16},
  month     = may,
  year      = 2019,
  language  = {en}
}

@article{Harville1992-vj,
  title     = {Mean Squared Error of Estimation or Prediction under a General
               Linear Model},
  author    = {Harville, David A and Jeske, Daniel R},
  journal   = {J. Am. Stat. Assoc.},
  publisher = {Taylor \& Francis},
  volume    = 87,
  number    = 419,
  pages     = {724--731},
  month     = sep,
  year      = 1992
}

@article{Ng2020-zc,
  title     = {The influence of training sample size on the accuracy of deep
               learning models for the prediction of soil properties with
               near-infrared spectroscopy data},
  author    = {Ng, Wartini and Minasny, Budiman and Mendes, Wanderson de Sousa
               and Dematt{\^e}, Jos{\'e} Alexandre Melo},
  journal   = {SOIL},
  publisher = {Copernicus GmbH},
  volume    = 6,
  number    = 2,
  pages     = {565--578},
  month     = nov,
  year      = 2020,
  language  = {en}
}

@article{Geng2021-ok,
  title    = {{COL1A1} is a prognostic biomarker and correlated with immune
              infiltrates in lung cancer},
  author   = {Geng, Qishun and Shen, Zhibo and Li, Lifeng and Zhao, Jie},
  journal  = {PeerJ},
  volume   = 9,
  pages    = {e11145},
  month    = mar,
  year     = 2021,
  language = {en}
}

@inproceedings{Gaonkar2016-ch,
  title      = {Deep learning in the small sample size setting: cascaded feed
                forward neural networks for medical image segmentation},
  booktitle  = {Medical Imaging 2016: {Computer-Aided} Diagnosis},
  author     = {Gaonkar, Bilwaj and Hovda, David and Martin, Neil and Macyszyn,
                Luke},
  publisher  = {SPIE},
  volume     = 9785,
  pages      = {646--653},
  month      = mar,
  year       = 2016,
  language   = {en},
  conference = {Medical Imaging 2016: Computer-Aided Diagnosis}
}

@article{Beatty2015-lv,
  title     = {Immune Escape Mechanisms as a Guide for Cancer Immunotherapy},
  author    = {Beatty, Gregory L and Gladney, Whitney L},
  journal   = {Clin. Cancer Res.},
  publisher = {American Association for Cancer Research},
  volume    = 21,
  number    = 4,
  pages     = {687--692},
  month     = feb,
  year      = 2015,
  language  = {en}
}

@misc{Ou2016-er,
  title        = {Asymmetric Transitivity Preserving Graph Embedding},
  booktitle    = {{ACM} Conferences},
  author       = {Ou, Mingdong and Cui, Peng and Pei, Jian and Zhang, Ziwei and
                  Zhu, Wenwu},
  month        = aug,
  year         = 2016,
  howpublished = {\url{https://dl.acm.org/doi/10.1145/2939672.2939751}},
  note         = {Accessed: 2022-9-12},
  language     = {en}
}

@article{Devlin2018-ag,
  title         = {{BERT}: Pre-training of Deep Bidirectional Transformers for
                   Language Understanding},
  author        = {Devlin, Jacob and Chang, Ming-Wei and Lee, Kenton and
                   Toutanova, Kristina},
  month         = oct,
  year          = 2018,
  archiveprefix = {arXiv},
  primaryclass  = {cs.CL},
  eprint        = {1810.04805}
}

@article{Gentleman2004-fg,
  title    = {Bioconductor: open software development for computational biology
              and bioinformatics},
  author   = {Gentleman, Robert C and Carey, Vincent J and Bates, Douglas M and
              Bolstad, Ben and Dettling, Marcel and Dudoit, Sandrine and Ellis,
              Byron and Gautier, Laurent and Ge, Yongchao and Gentry, Jeff and
              Hornik, Kurt and Hothorn, Torsten and Huber, Wolfgang and Iacus,
              Stefano and Irizarry, Rafael and Leisch, Friedrich and Li, Cheng
              and Maechler, Martin and Rossini, Anthony J and Sawitzki, Gunther
              and Smith, Colin and Smyth, Gordon and Tierney, Luke and Yang,
              Jean Y H and Zhang, Jianhua},
  journal  = {Genome Biol.},
  volume   = 5,
  number   = 10,
  pages    = {R80},
  month    = sep,
  year     = 2004,
  language = {en}
}

@article{Wang2022-ev,
  title     = {Cross-modal fusion for multi-label image classification with
               attention mechanism},
  author    = {Wang, Yangtao and Xie, Yanzhao and Zeng, Jiangfeng and Wang,
               Hanpin and Fan, Lisheng and Song, Yufan},
  journal   = {Comput. Electr. Eng.},
  publisher = {Elsevier},
  volume    = 101,
  pages     = {108002},
  month     = jul,
  year      = 2022
}

@article{Grover2016-bo,
  title     = {node2vec: Scalable Feature Learning for Networks},
  author    = {Grover, Aditya and Leskovec, Jure},
  journal   = {KDD},
  publisher = {dl.acm.org},
  volume    = 2016,
  pages     = {855--864},
  month     = aug,
  year      = 2016,
  language  = {en}
}

@article{Van_Dam2019-cf,
  title    = {{RANK-RANKL} Signaling in Cancer of the Uterine Cervix: A Review},
  author   = {van Dam, Peter A and Verhoeven, Yannick and Jacobs, Julie and
              Wouters, An and Tjalma, Wiebren and Lardon, Filip and Van den
              Wyngaert, Tim and Dewulf, Jonatan and Smits, Evelien and
              Colpaert, C{\'e}cile and Prenen, Hans and Peeters, Marc and
              Lammens, Martin and Trinh, Xuan Bich},
  journal  = {Int. J. Mol. Sci.},
  volume   = 20,
  number   = 9,
  month    = may,
  year     = 2019,
  language = {en}
}

@article{Iuchi2021-uh,
  title     = {Representation learning applications in biological sequence
               analysis},
  author    = {Iuchi, Hitoshi and Matsutani, Taro and Yamada, Keisuke and
               Iwano, Natsuki and Sumi, Shunsuke and Hosoda, Shion and Zhao,
               Shitao and Fukunaga, Tsukasa and Hamada, Michiaki},
  journal   = {Comput. Struct. Biotechnol. J.},
  publisher = {Elsevier},
  volume    = 19,
  pages     = {3198--3208},
  month     = jan,
  year      = 2021
}

@article{Kulmanov2022-jk,
  title    = {{DeepGOZero}: improving protein function prediction from sequence
              and zero-shot learning based on ontology axioms},
  author   = {Kulmanov, Maxat and Hoehndorf, Robert},
  journal  = {Bioinformatics},
  volume   = 38,
  number   = {Suppl 1},
  pages    = {i238--i245},
  month    = jun,
  year     = 2022,
  language = {en}
}

@article{Peiran2021-af,
  title     = {{FSL-Kla}: A few-shot learning-based multi-feature hybrid system
               for lactylation site prediction},
  author    = {Peiran, Jiang and Wanshan, Ning and Yunshu, Shi and Saijun, Mo and
               Chuan, Liu and Haoran, Zhou and Kangdong, Liu and Yaping, Guo},
  journal   = {Comput. Struct. Biotechnol. J.},
  publisher = {Elsevier},
  volume    = 19,
  pages     = {4497--4509},
  month     = jan,
  year      = 2021
}

@article{Li2021-ll,
  title    = {De novo Prediction of Moonlighting Proteins Using Multimodal Deep
              Ensemble Learning},
  author   = {Li, Ying and Zhao, Jianing and Liu, Zhaoqian and Wang, Cankun and
              Wei, Lizheng and Han, Siyu and Du, Wei},
  journal  = {Front. Genet.},
  volume   = 12,
  pages    = {630379},
  month    = mar,
  year     = 2021,
  language = {en}
}

@article{Akbar2022-cs,
  title    = {In silico proof of principle of machine learning-based antibody
              design at unconstrained scale},
  author   = {Akbar, Rahmad and Robert, Philippe A and Weber, C{\'e}dric R and
              Widrich, Michael and Frank, Robert and Pavlovi{\'c}, Milena and
              Scheffer, Lonneke and Chernigovskaya, Maria and Snapkov, Igor and
              Slabodkin, Andrei and Mehta, Brij Bhushan and Miho, Enkelejda and
              Lund-Johansen, Fridtjof and Andersen, Jan Terje and Hochreiter,
              Sepp and Hob{\ae}k Haff, Ingrid and Klambauer, G{\"u}nter and
              Sandve, Geir Kjetil and Greiff, Victor},
  journal  = {MAbs},
  volume   = 14,
  number   = 1,
  pages    = {2031482},
  month    = jan,
  year     = 2022,
  language = {en}
}

@article{Elnaggar2022-qd,
  title     = {{ProtTrans}: Toward Understanding the Language of Life Through
               {Self-Supervised} Learning},
  author    = {Elnaggar, A and Heinzinger, M and Dallago, C and Rehawi, G and
               Wang, Y and Jones, L and Gibbs, T and Feher, T and Angerer, C
               and Steinegger, M and Bhowmik, D and Rost, B},
  journal   = {IEEE Trans. Pattern Anal. Mach. Intell.},
  publisher = {IEEE Trans Pattern Anal Mach Intell},
  volume    = 44,
  number    = 10,
  month     = oct,
  year      = 2022
}

@article{Jayakumar2020-va,
  title  = {Multiplicative interactions and where to find them},
  author = {Jayakumar, Siddhant M and Czarnecki, Wojciech M and Menick, Jacob
            and Schwarz, Jonathan and Rae, Jack and Osindero, Simon and Teh,
            Yee Whye and Harley, Tim and Pascanu, Razvan},
  year   = 2020
}

@article{Giri2021-ch,
  title    = {{MultiPredGO}: Deep {Multi-Modal} Protein Function Prediction by
              Amalgamating Protein Structure, Sequence, and Interaction
              Information},
  author   = {Giri, Swagarika Jaharlal and Dutta, Pratik and Halani, Parth and
              Saha, Sriparna},
  journal  = {IEEE J Biomed Health Inform},
  volume   = 25,
  number   = 5,
  pages    = {1832--1838},
  month    = may,
  year     = 2021,
  language = {en}
}

@article{He2022-cb,
  title     = {Pentose Phosphate Pathway Regulates Tolerogenic Apoptotic Cell
               Clearance and Immune Tolerance},
  author    = {He, Dan and Mao, Qiangdongzi and Jia, Jialin and Wang, Zhiyu and
               Liu, Yu and Liu, Tingting and Luo, Bangwei and Zhang, Zhiren},
  journal   = {Front. Immunol.},
  publisher = {Frontiers},
  volume    = 12,
  month     = jan,
  year      = 2022,
  language  = {en}
}

@article{Zhou2019-iw,
  title     = {The {CAFA} challenge reports improved protein function
               prediction and new functional annotations for hundreds of genes
               through experimental screens},
  author    = {Zhou, Naihui and Jiang, Yuxiang and Bergquist, Timothy R and
               Lee, Alexandra J and Kacsoh, Balint Z and Crocker, Alex W and
               Lewis, Kimberley A and Georghiou, George and Nguyen, Huy N and
               Hamid, Md Nafiz and Davis, Larry and Dogan, Tunca and Atalay,
               Volkan and Rifaioglu, Ahmet S and Dalk{\i}ran, Alperen and Cetin
               Atalay, Rengul and Zhang, Chengxin and Hurto, Rebecca L and
               Freddolino, Peter L and Zhang, Yang and Bhat, Prajwal and Supek,
               Fran and Fern{\'a}ndez, Jos{\'e} M and Gemovic, Branislava and
               Perovic, Vladimir R and Davidovi{\'c}, Radoslav S and Sumonja,
               Neven and Veljkovic, Nevena and Asgari, Ehsaneddin and Mofrad,
               Mohammad R K and Profiti, Giuseppe and Savojardo, Castrense and
               Martelli, Pier Luigi and Casadio, Rita and Boecker, Florian and
               Schoof, Heiko and Kahanda, Indika and Thurlby, Natalie and
               McHardy, Alice C and Renaux, Alexandre and Saidi, Rabie and
               Gough, Julian and Freitas, Alex A and Antczak, Magdalena and
               Fabris, Fabio and Wass, Mark N and Hou, Jie and Cheng, Jianlin
               and Wang, Zheng and Romero, Alfonso E and Paccanaro, Alberto and
               Yang, Haixuan and Goldberg, Tatyana and Zhao, Chenguang and
               Holm, Liisa and T{\"o}r{\"o}nen, Petri and Medlar, Alan J and
               Zosa, Elaine and Borukhov, Itamar and Novikov, Ilya and Wilkins,
               Angela and Lichtarge, Olivier and Chi, Po-Han and Tseng,
               Wei-Cheng and Linial, Michal and Rose, Peter W and Dessimoz,
               Christophe and Vidulin, Vedrana and Dzeroski, Saso and Sillitoe,
               Ian and Das, Sayoni and Lees, Jonathan Gill and Jones, David T
               and Wan, Cen and Cozzetto, Domenico and Fa, Rui and Torres,
               Mateo and Warwick Vesztrocy, Alex and Rodriguez, Jose Manuel and
               Tress, Michael L and Frasca, Marco and Notaro, Marco and Grossi,
               Giuliano and Petrini, Alessandro and Re, Matteo and Valentini,
               Giorgio and Mesiti, Marco and Roche, Daniel B and Reeb, Jonas
               and Ritchie, David W and Aridhi, Sabeur and Alborzi, Seyed
               Ziaeddin and Devignes, Marie-Dominique and Koo, Da Chen Emily
               and Bonneau, Richard and Gligorijevi{\'c}, Vladimir and Barot,
               Meet and Fang, Hai and Toppo, Stefano and Lavezzo, Enrico and
               Falda, Marco and Berselli, Michele and Tosatto, Silvio C E and
               Carraro, Marco and Piovesan, Damiano and Ur Rehman, Hafeez and
               Mao, Qizhong and Zhang, Shanshan and Vucetic, Slobodan and
               Black, Gage S and Jo, Dane and Suh, Erica and Dayton, Jonathan B
               and Larsen, Dallas J and Omdahl, Ashton R and McGuffin, Liam J
               and Brackenridge, Danielle A and Babbitt, Patricia C and Yunes,
               Jeffrey M and Fontana, Paolo and Zhang, Feng and Zhu, Shanfeng
               and You, Ronghui and Zhang, Zihan and Dai, Suyang and Yao,
               Shuwei and Tian, Weidong and Cao, Renzhi and Chandler, Caleb and
               Amezola, Miguel and Johnson, Devon and Chang, Jia-Ming and Liao,
               Wen-Hung and Liu, Yi-Wei and Pascarelli, Stefano and Frank,
               Yotam and Hoehndorf, Robert and Kulmanov, Maxat and Boudellioua,
               Imane and Politano, Gianfranco and Di Carlo, Stefano and Benso,
               Alfredo and Hakala, Kai and Ginter, Filip and Mehryary, Farrokh
               and Kaewphan, Suwisa and Bj{\"o}rne, Jari and Moen, Hans and
               Tolvanen, Martti E E and Salakoski, Tapio and Kihara, Daisuke
               and Jain, Aashish and {\v S}muc, Tomislav and Altenhoff, Adrian
               and Ben-Hur, Asa and Rost, Burkhard and Brenner, Steven E and
               Orengo, Christine A and Jeffery, Constance J and Bosco, Giovanni
               and Hogan, Deborah A and Martin, Maria J and O'Donovan, Claire
               and Mooney, Sean D and Greene, Casey S and Radivojac, Predrag
               and Friedberg, Iddo},
  journal   = {Genome Biol.},
  publisher = {BioMed Central},
  volume    = 20,
  number    = 1,
  pages     = {1--23},
  month     = nov,
  year      = 2019,
  language  = {en}
}

@article{Hirschmeier2020-lc,
  title    = {Improving Recall and Precision in Unsupervised {Multi-Label}
              Document Classification Tasks by Combining Word Embeddings with
              {TF-IDF}},
  author   = {Hirschmeier, Stefan and Melsbach, Johannes Werner and Schoder,
              Detlef and Stahlmann, Sven},
  series   = {ECIS 2020 Research Papers},
  year     = 2020
}

@article{Ahmed2022-nj,
  title    = {Glutathione peroxidase 2 is a metabolic driver of the tumor
              immune microenvironment and immune checkpoint inhibitor response},
  author   = {Ahmed, Kazi Mokim and Veeramachaneni, Ratna and Deng, Defeng and
              Putluri, Nagireddy and Putluri, Vasanta and Cardenas, Maria F and
              Wheeler, David A and Decker, William K and Frederick, Andy I and
              Kazi, Sawad and Sikora, Andrew G and Sandulache, Vlad C and
              Frederick, Mitchell J},
  journal  = {J Immunother Cancer},
  volume   = 10,
  number   = 8,
  month    = aug,
  year     = 2022,
  language = {en}
}

@misc{Chen2019-qo,
  title   = {{BioSentVec}: creating sentence embeddings for biomedical texts},
  author  = {Chen, Qingyu and Peng, Yifan and Lu, Zhiyong},
  journal = {2019 IEEE International Conference on Healthcare Informatics
             (ICHI)},
  year    = 2019
}

@article{Zhong2022-yg,
  title    = {Long-distance dependency combined multi-hop graph neural networks
              for protein-protein interactions prediction},
  author   = {Zhong, Wen and He, Changxiang and Xiao, Chen and Liu, Yuru and
              Qin, Xiaofei and Yu, Zhensheng},
  journal  = {BMC Bioinformatics},
  volume   = 23,
  number   = 1,
  pages    = {521},
  month    = dec,
  year     = 2022,
  language = {en}
}

@article{Chen_undated-iz,
  title   = {edgeR: Empirical analysis of digital gene expression data in {R}},
  author  = {Chen, Y and Lun, A and McCarthy, D and Ritchie, M E and Phipson, B
             and {others}},
  journal = {Version: Release (3.12)}
}

@article{Fridman2003-pa,
  title    = {Cell surface association of matrix metalloproteinase-9
              (gelatinase B)},
  author   = {Fridman, Rafael and Toth, Marta and Chvyrkova, Irina and Meroueh,
              Samy O and Mobashery, Shahriar},
  journal  = {Cancer Metastasis Rev.},
  volume   = 22,
  number   = {2-3},
  pages    = {153--166},
  year     = 2003,
  language = {en}
}

@article{Ohue2019-tf,
  title    = {Regulatory {T} (Treg) cells in cancer: Can Treg cells be a new
              therapeutic target?},
  author   = {Ohue, Yoshihiro and Nishikawa, Hiroyoshi},
  journal  = {Cancer Sci.},
  volume   = 110,
  number   = 7,
  pages    = {2080--2089},
  month    = jul,
  year     = 2019,
  language = {en}
}

@article{Yu2010-nc,
  title    = {{GOSemSim}: an {R} package for measuring semantic similarity
              among {GO} terms and gene products},
  author   = {Yu, Guangchuang and Li, Fei and Qin, Yide and Bo, Xiaochen and
              Wu, Yibo and Wang, Shengqi},
  journal  = {Bioinformatics},
  volume   = 26,
  number   = 7,
  pages    = {976--978},
  month    = apr,
  year     = 2010,
  language = {en}
}

@article{Wang2018-qo,
  title         = {{GLUE}: A {Multi-Task} Benchmark and Analysis Platform for
                   Natural Language Understanding},
  author        = {Wang, Alex and Singh, Amanpreet and Michael, Julian and
                   Hill, Felix and Levy, Omer and Bowman, Samuel R},
  month         = apr,
  year          = 2018,
  archiveprefix = {arXiv},
  primaryclass  = {cs.CL},
  eprint        = {1804.07461}
}

@article{Littmann2021-oj,
  title     = {Embeddings from deep learning transfer {GO} annotations beyond
               homology},
  author    = {Littmann, Maria and Heinzinger, Michael and Dallago, Christian
               and Olenyi, Tobias and Rost, Burkhard},
  journal   = {Sci. Rep.},
  publisher = {Nature Publishing Group},
  volume    = 11,
  number    = 1,
  pages     = {1--14},
  month     = jan,
  year      = 2021,
  language  = {en}
}

@article{You2021-fk,
  title     = {{DeepGraphGO}: graph neural network for large-scale,
               multispecies protein function prediction},
  author    = {You, Ronghui and Yao, Shuwei and Mamitsuka, Hiroshi and Zhu,
               Shanfeng},
  journal   = {Bioinformatics},
  publisher = {Oxford Academic},
  volume    = 37,
  number    = {Supplement\_1},
  pages     = {i262--i271},
  month     = jul,
  year      = 2021,
  language  = {en}
}

@article{Gryshchenko2021-he,
  title    = {Calcium Signaling in Pancreatic Immune Cells},
  author   = {Gryshchenko, Oleksiy and Gerasimenko, Julia V and Petersen, Ole H
              and Gerasimenko, Oleg V},
  journal  = {Function (Oxf)},
  volume   = 2,
  number   = 1,
  pages    = {zqaa026},
  year     = 2021,
  language = {en}
}

@misc{noauthor_2015-kr,
  title        = {Cancer mortality for common cancers},
  booktitle    = {Cancer Research {UK}},
  month        = may,
  year         = 2015,
  howpublished = {\url{https://www.cancerresearchuk.org/health-professional/cancer-statistics/mortality/common-cancers-compared}},
  note         = {Accessed: 2023-3-18},
  language     = {en}
}

@article{He2018-ep,
  title    = {Large-scale prediction of protein ubiquitination sites using a
              multimodal deep architecture},
  author   = {He, Fei and Wang, Rui and Li, Jiagen and Bao, Lingling and Xu,
              Dong and Zhao, Xiaowei},
  journal  = {BMC Syst. Biol.},
  volume   = 12,
  number   = {Suppl 6},
  pages    = {109},
  month    = nov,
  year     = 2018,
  language = {en}
}

@article{Wittmann2021-ka,
  title     = {Informed training set design enables efficient machine
               learning-assisted directed protein evolution},
  author    = {Wittmann, Bruce J and Yue, Yisong and Arnold, Frances H},
  journal   = {Cell Systems},
  publisher = {Cell Press},
  volume    = 12,
  number    = 11,
  pages     = {1026--1045.e7},
  month     = nov,
  year      = 2021
}

@misc{Moghadasi2020-se,
  title   = {{Sent2Vec}: A New Sentence Embedding Representation With
             Sentimental Semantic},
  author  = {Moghadasi, Mahdi Naser and Zhuang, Yu},
  journal = {2020 IEEE International Conference on Big Data (Big Data)},
  year    = 2020
}

@incollection{Imambi2021-ym,
  title     = {{PyTorch}},
  booktitle = {Programming with {TensorFlow}: Solution for Edge Computing
               Applications},
  author    = {Imambi, Sagar and Prakash, Kolla Bhanu and Kanagachidambaresan,
               G R},
  editor    = {Prakash, Kolla Bhanu and Kanagachidambaresan, G R},
  publisher = {Springer International Publishing},
  pages     = {87--104},
  year      = 2021,
  address   = {Cham}
}

@article{Cui2020-ye,
  title     = {{MV-RNN}: A {Multi-View} Recurrent Neural Network for Sequential
               Recommendation},
  author    = {Cui, Qiang and Wu, Shu and Liu, Qiang and Zhong, Wen and Wang,
               Liang},
  journal   = {IEEE Trans. Knowl. Data Eng.},
  publisher = {IEEE Educational Activities Department},
  volume    = 32,
  number    = 2,
  pages     = {317--331},
  month     = feb,
  year      = 2020,
  address   = {USA}
}

@article{Brandau2011,
  author = {Brandau, S. and Trellakis, S. and Bruderek, K. and Schmaltz, D. and Steller, G. and Elian, M. and Suttmann, H. and Schenck, M. and Welling, J. and Zabel, P. and Lang, S.},
  title = {Myeloid-derived suppressor cells in the peripheral blood of cancer patients contain a subset of immature neutrophils with impaired migratory properties},
  journal = {Journal of Leukocyte Biology},
  volume = {89},
  number = {2},
  pages = {311--317},
  year = {2011},
  doi = {10.1189/jlb.0310162}
}

@article{Jadus2012,
  author = {Martin R. Jadus and Josephine Natividad and Anthony Mai and Yi Ouyang and Nils Lambrecht and Sandor Szabo and Lisheng Ge and Neil Hoa and Maria G. Dacosta-Iyer},
  title = {Lung Cancer: A Classic Example of Tumor Escape and Progression While Providing Opportunities for Immunological Intervention},
  journal = {Clinical and Developmental Immunology},
  volume = {2012},
  pages = {160724},
  year = {2012},
  doi = {10.1155/2012/160724}
}

@article{Hunn2016,
  author = {Hunn, D. and Bronte, V.},
  title = {Immune suppressive mechanisms in the tumor microenvironment},
  journal = {Current Opinion in Immunology},
  volume = {39},
  pages = {1--6},
  year = {2016},
  doi = {10.1016/j.coi.2015.10.009}
}

@article{Vinay2015,
  author = {Dass S. Vinay and Elizabeth P. Ryan and Graham Pawelec and Wamidh H. Talib and John Stagg and Eyad Elkord and Terry Lichtor and William K. Decker and Richard L. Whelan and H. M. C. Shantha Kumara and Emanuela Signori and Kanya Honoki and Alexandros G. Georgakilas and Amr Amin and William G. Helferich and Chandra S. Boosani and Gunjan Guha and Maria Rosa Ciriolo and Sophie Chen and Sulma I. Mohammed and Asfar S. Azmi and W. Nicol Keith and Alan Bilsland and Dipita Bhakta and Dorota Halicka and Hiromasa Fujii and Katia Aquilano and S. Salman Ashraf and Somaira Nowsheen and Xujuan Yang and Beom K. Choi and Byoung S. Kwon},
  title = {Immune evasion in cancer: Mechanistic basis and therapeutic strategies},
  journal = {Seminars in Cancer Biology},
  volume = {35},
  pages = {S185--S198},
  year = {2015},
  doi = {10.1016/j.semcancer.2015.03.004}
}

@article{Leone2018,
  author = {Leone, K. and Poggiana, C. and Zamarchi, R.},
  title = {The Interplay between Circulating Tumor Cells and the Immune System: From Immune Escape to Cancer Immunotherapy},
  journal = {Diagnostics},
  volume = {8},
  number = {3},
  pages = {59},
  year = {2018},
  doi = {10.3390/diagnostics8030059}
}

@article{Huang2017,
  author = {Renxiang Huang and Dongyang Zhang and Feng Li and Zili Xiao and Meiling Wu and Dongyun Shi and Ping Xiang and Zhijun Bao},
  title = {Loss of Fas expression and high expression of HLA-E promoting the immune escape of early colorectal cancer cells},
  journal = {Oncology Letters},
  volume = {13},
  pages = {3379--3386},
  year = {2017},
  doi = {10.3892/ol.2017.5921}
}

@article{Carosella2015,
  author = {Carosella, E. and Ploussard, G. and LeMaoult, J. and Desgrandchamps, F.},
  title = {Systematic Review of Immunotherapy in Urologic Cancer: Evolving Roles for Targeting of CTLA-4, PD-1/PD-L1, and HLA-G},
  journal = {European Urology},
  volume = {68},
  number = {2},
  pages = {267--279},
  year = {2015},
  doi = {10.1016/j.eururo.2015.02.040}
}

@article{Zelenay2015,
  author = {Zelenay, S. and van der Veen, A. G. and Böttcher, J. P. and Snelgrove, K. J. and Rogers, N. and Acton, S. E. and Chakravarty, P. and Girotti, M. R. and Marais, R. and Quezada, S. A. and Sahai, E. and Reis e Sousa, C.},
  title = {Cyclooxygenase-Dependent Tumor Growth through Evasion of Immunity},
  journal = {Cell},
  volume = {162},
  pages = {1257--1270},
  year = {2015},
  doi = {10.1016/j.cell.2015.08.015}
}

@article{Kotteas2014,
  author = {Kotteas, E. A. and Boulas, P. and Gkiozos, I. and Tsagkouli, S. and Tsoukalas, G. and Syrigos, K. N.},
  title = {The Intercellular Cell Adhesion Molecule-1 (ICAM-1) in Lung Cancer: Implications for Disease Progression and Prognosis},
  journal = {Anticancer Research},
  volume = {34},
  number = {9},
  pages = {4665--4672},
  year = {2014}
}

@article{Suzek2015,
  author = {Suzek, B. E. and Wang, Y. and Huang, H. and McGarvey, P. B. and Wu, C. H. and {The UniProt Consortium}},
  title = {UniRef clusters: a comprehensive and scalable alternative for improving sequence similarity searches},
  journal = {Bioinformatics},
  volume = {31},
  number = {6},
  pages = {926--932},
  year = {2015},
  doi = {10.1093/bioinformatics/btu739}
}
\end{document}